\documentclass[11pt,a4paper]{article}

\usepackage[T1]{fontenc}
\usepackage[utf8]{inputenc}
\usepackage{lmodern}
\usepackage[margin=2.4cm]{geometry}
\usepackage{amsmath,amssymb,amsthm,bm}
\usepackage{graphicx,tikz}
\usepackage{booktabs}
\usepackage[numbers,sort&compress]{natbib}
\usepackage[hidelinks]{hyperref}
\graphicspath{{figs/}}

\newcommand{\dd}{\mathrm{d}}
\newcommand{\E}{\mathbb{E}}
\newcommand{\Var}{\mathrm{Var}}
\newcommand{\vx}{\bm{x}}
\newcommand{\vn}{\bm{n}}
\newcommand{\Lop}{\mathcal{L}^{*}}
\newcommand{\nub}{\bar\nu}

\title{Estimating subcritical multiplication from stochastic neutron noise: a branching Brownian motion model for leakage counts}
\author{Antoine Bessemoulin$^{1,2}$ and Eric Dumonteil$^1$\\[4pt]
\small $^1$Université Paris-Saclay, CEA, IRFU, Département de Physique Nucléaire,\\
\small 91191 Gif-sur-Yvette, France\\
\small $^2$CMAP, CNRS, École polytechnique, Institut Polytechnique de Paris,\\
\small 91120 Palaiseau, France}
\date{November 19, 2026}

\begin{document}
\maketitle

\begin{abstract}
A neutron detector placed outside a subcritical system records fission
chains as correlated bursts of counts. The time dependence of these
correlations gives access to the $\alpha$ eigenvalues governing the decay of
the system and, with additional physical information, to its effective
multiplication factor $k_{\mathrm{eff}}$. The usual analysis relies on point
kinetics, which ignores where chains are born and which neutrons escape.
We examine this approximation in a model simple enough to be solved:
branching Brownian motion, that is, one-group neutron diffusion with capture
and fission, in a domain whose absorbing boundary plays the role of the
detector. For a stationary Poisson source we derive the Feynman $Y$ function
(the count variance-to-mean ratio minus one) by following pairs of neutrons
from their common ancestor to their detection. In a homogeneous cube the
result is an explicit sum over spatial modes whose weights are fixed by the
geometry and the source. We compare it with direct branching simulations and
with one-speed OpenMC transport calculations in which fission is sampled
with analog reaction selection and FREYA multiplicities. At the longest simulated gate the model agrees
with transport to about $1\%$ at $k=0.8$ and $0.9$, and to $10\%$ at
$k=0.5$. The corresponding point-model discrepancies range from $25$ to $48\%$. When geometry, source
and material parameters are known, retaining the spatial modes also reduces
the bias on the inferred multiplication factor. Two extensions prepare the
use of the model in inverse problems. Correlations between opposite faces
distinguish systems that share the same fundamental decay rate but differ in
leakage, and a hierarchy of three stationary problems links this spatial
benchmark to multiplicity counting. Modal convergence, detector imbalance
and spectral evolution remain limitations for quantitative inference.
\end{abstract}

\section{Introduction}

In various operational contexts, neutron detectors can help monitor how closely a system approaches
criticality. During reactor startup, the approach is deliberate and must be
followed. In spent fuel pools and in fuel-cycle facilities that handle
fissile material, the margin to criticality must be maintained. A related
concern arises during fuel debris retrieval at Fukushima Daiichi, where
changes in the arrangement of fuel and water could affect the risk of
recriticality~\citep{jaeaDebris}.

One observable that carries this information is the spontaneous fluctuation
of the neutron population as fission chains grow and die out, known as
\emph{stochastic neutron noise} (historically, \emph{pile noise}). Neutrons
of the same chain share a common ancestor, so their arrivals at a detector
are correlated. The time dependence of these correlations is set by the
$\alpha$ eigenvalues that govern the decay of the neutron population. With
additional information, in particular the neutron generation time, it also
yields an estimate of $k_{\mathrm{eff}}$. This is the basis of the
Feynman-$\alpha$, Rossi-$\alpha$ and factorial-moment
methods~\citep{feynman1956dispersion,hage1985correlation,cifarelli1986models,pazsit2007neutron}.
In their simplest form these methods rest on point kinetics: every neutron
is assumed to have the same chance of inducing a fission, of being captured
and of being detected, wherever it is.

Space matters for two reasons. First, common ancestry clusters neutrons in
space as well as in
time~\citep{dumonteil2014,zoia2014clustering,dumonteil2021patchy}. Second, a
detector outside the system counts only the neutrons that escape, and a pair
of correlated neutrons born near the centre does not have the same chance of
being counted as a pair born near the surface. Fuel debris retrieval gives a
concrete illustration. Moving or collecting fragments changes the escape
paths of neutrons, while changes in water content and composition affect
their slowing down, capture and fission. JAEA lists subcriticality
measurement and neutron monitoring among the open issues for retrieval and
for debris collected in water-system components~\citep{jaeaMonitoring2026}.
A count rate, or a single decay constant, need not distinguish a change in
leakage from a change in multiplication.

In this paper we study these effects in a model that can be solved
explicitly. Neutrons are represented by a branching Brownian motion (BBM):
particles diffuse, are captured or undergo fission, and are counted when
they reach an absorbing boundary. This is one-group neutron transport in the
diffusion limit. The geometry is a homogeneous cube, in which leakage and
multiplication are each controlled by a single coefficient: a leakage scale
$a$ and a net growth rate $\rho$. The cube is a test of what can be inferred
from spatially resolved noise. It is not a representation of a debris
configuration, nor a qualified criticality-monitoring system. The paper
makes four contributions.
\begin{enumerate}
\item A compact expression for the Feynman $Y$ function of boundary counts
  in an arbitrary bounded domain, and its explicit modal form in a cube
  (Sec.~\ref{sec:theory}). It is checked against direct BBM simulations
  (Sec.~\ref{sec:bbm-mc}).
\item An analog fission option in OpenMC, coupled to FREYA, so that
  transport simulations preserve the pair correlations created by fission
  (Sec.~\ref{sec:analog}).
\item A comparison with one-speed and continuous-energy transport that
  quantifies the bias of point kinetics, on $Y$ and on the multiplication
  factor inferred from it (Sec.~\ref{sec:openmc}).
\item Two spatial benchmarks for inverse problems: opposite-face
  correlations that separate leakage from net multiplication, and a
  three-moment hierarchy that extends point-model multiplicity counting
  (Sec.~\ref{sec:inversebenchmark}).
\end{enumerate}

\paragraph{Relation to previous work.}
Spatial corrections to neutron counting are well established, and we do not
propose a first spatial noise theory. Mu\~noz-Cobo
et al.~\citep{munoz2011} obtained multimode Feynman and Rossi expressions.
Yamamoto studied spatial and energy higher modes and compared them with
Monte Carlo calculations~\citep{yamamoto2011modes,yamamoto2013energy}. The
relation between the Feynman and Rossi observables holds beyond point
kinetics~\citep{degweker2016}. Stationary spatial plateaus and the numerical
validation of counting methods are treated in
Refs.~\citep{gabrieli2023,guillevic2025}, and higher-order spatial
correlations were derived by Endo et al.~\citep{endo2006third}. Within this
body of theory we consider one specific problem, counting at the boundary,
and seek a benchmark that is tractable and in which the origin of each term
stays explicit. We build on the Green's-function description of confined
branching Brownian motions~\citep{zoia2014clustering}. That work observes
the population inside the domain at a given time; we observe its exits
through the boundary during a counting gate. The stationary hierarchy of
Sec.~\ref{sec:boehnel} is the boundary-counting counterpart of B\"ohnel's
point-model construction~\citep{boehnel1985,prasad2012}. It can serve as a
forward model for inverse problems, intermediate between point kinetics and
full transport, in the approaches discussed by
Houpert~\citep{houpert2024slides} and Lartaud et al.~\citep{lartaud2025}.

A numerical comparison is meaningful only if the transport code preserves
the correlations created by fission. Petit and
Dumonteil~\citep{petit2015analog} implemented fully analog neutron transport
in TRIPOLI-4~\citep{brun2015tripoli} for nuclear instrumentation. It was the
transport basis of the TRIPOLI-4--FREYA work of Verbeke and
Petit~\citep{verbeke2016stochastic}, who applied correlated fission sampling
to neutron multiplicity counting and Bayesian uncertainty quantification.
FREYA supplies event-by-event fission multiplicities and outgoing
particles~\citep{verbeke2018correlated,vogt2009freya}. More recently, PyNMC
coupled OpenMC to FREYA for spontaneous and induced
fission~\citep{fichtlscherer2026}. These works establish the importance of
analog transport when particle histories are recorded directly. Non-analog
noise estimators are also possible~\citep{yamamoto2011nonanalog}.

\paragraph{Outline.}
Section~\ref{sec:theory} derives the counting statistics of the BBM and
Sec.~\ref{sec:bbm-mc} checks them against direct simulation.
Section~\ref{sec:analog} describes the analog fission treatment in OpenMC.
Section~\ref{sec:openmc} compares the model with transport and examines the
inference of a decay rate and a multiplication factor.
Section~\ref{sec:inversebenchmark} presents the two benchmarks for inverse
problems, and Sec.~\ref{sec:conclusion} discusses the limits of the
approach. Table~\ref{tab:notation} collects the main notation.

\section{Counting statistics of a branching Brownian motion}
\label{sec:theory}

\subsection{Model}

Consider particles evolving independently in a bounded domain
$V\subset\mathbb{R}^d$ with absorbing boundary $\partial V$
(Fig.~\ref{fig:benchmark}). Following Ref.~\citep{zoia2014clustering}, each
particle performs a Brownian motion with diffusion coefficient $D$. At rate
$\lambda$ it undergoes a reproduction event, which replaces it by $j$
particles with probability $q_j$; $j=0$ is a capture.
We write $G(z)=\sum_jq_jz^j$ for the offspring generating function, and
$\nu_1=G'(1)$, $\nu_2=G''(1)<\infty$ for its first two factorial moments:
the mean number of offspring, and the mean number of ordered pairs of
offspring. The coefficients are uniform and time independent. Particles
interact with the medium but not with one another.

A particle that reaches $\partial V$ leaks out, and is recorded by the
detector with probability $\varepsilon$. The efficiency
$0<\varepsilon\le1$ is constant, each particle produces at most one count,
and there is no dead time or detector delay. This ideal detector separates
the effect of propagation inside the medium from that of the measuring
apparatus.

For one-speed neutrons of speed $v$ in the diffusion limit,
$D=v/(3\Sigma_t)$ for isotropic scattering. Reproduction events are
absorptions, $\lambda=v\Sigma_a$: a capture leaves no offspring and a
fission leaves $\nu$ neutrons, with probability $P(\nu)$. Hence
\begin{equation}
  q_0 = \frac{\Sigma_c}{\Sigma_a}+\frac{\Sigma_f}{\Sigma_a}P(0),\qquad
  q_j = \frac{\Sigma_f}{\Sigma_a}P(j)\ (j\ge1),
  \qquad\Rightarrow\qquad
  \lambda\nu_1 = v\nub\Sigma_f,\quad \lambda\nu_2 = v\Sigma_f\langle\nu(\nu-1)\rangle.
  \label{eq:mapping}
\end{equation}
Two combinations of these parameters control the first two counting moments. The
\emph{net growth rate} $\rho=\lambda(\nu_1-1)=v(\nub\Sigma_f-\Sigma_a)$ is
the rate at which the mean population would grow in an infinite medium. The
\emph{pair-production rate} $\lambda\nu_2$ is the rate at which a particle
creates pairs of correlated offspring. Scattering is represented by diffusion. Adding events that leave one
particle at the same position would not change these two counting moments,
provided $\rho$ and $\lambda\nu_2$ remain unchanged. Here $\lambda$ retains
its absorption-rate definition, including when converting decay rates to $k$.

\begin{table}[t]
\centering
\caption{Main notation.}
\label{tab:notation}
\small
\begin{tabular}{ll}
\toprule
$D$, $\lambda$ & diffusion coefficient; reproduction (absorption) rate \\
$q_j$, $G(z)$ & offspring distribution and its generating function \\
$\nu_1$, $\nu_2$, $\nu_3$ & factorial moments $G'(1)$, $G''(1)$, $G'''(1)$ \\
$\rho=\lambda(\nu_1-1)$ & net growth rate \\
$\lambda\nu_2$ & pair-production rate \\
$S$, $p(\vx)$ & source rate; normalized source density \\
$\varepsilon$ & detection probability of a leaking particle \\
$T$, $C$ & gate duration; number of counts in the gate \\
$Y(T)$, $Y_\infty$ & Feynman function; its long-gate limit \\
$h(\vx)$ & mean number of leaks of a chain started at $\vx$ \\
$\widehat c(\vx)$ & stationary particle density per unit source rate \\
$L$ & half-side of the diffusion cube \\
$a=D(\pi/2L)^2$ & leakage scale \\
$b_{\vn}=a|\vn|^2$, $\gamma_{\vn}=b_{\vn}-\rho$ & leakage rate and decay constant of mode $\vn$ \\
$b_1$, $\gamma_1$ & the same for the fundamental mode $\vn=(1,1,1)$ \\
$k=\lambda\nu_1/(\lambda+b_1)$ & one-group multiplication factor \\
$\Lambda=1/(\lambda\nu_1)$ & prompt generation time \\
$\kappa\simeq1.262$ & form factor of the fundamental mode, Eq.~\eqref{eq:Yfund} \\
\bottomrule
\end{tabular}
\end{table}

\begin{figure}[t]
\centering
\resizebox{\linewidth}{!}{
\begin{tikzpicture}[x=1cm,y=1cm,font=\fontsize{9}{11}\selectfont,>=stealth,
 box/.style={draw=black!35,rounded corners=2pt,fill=black!2,align=left,inner sep=7pt}]
\path (0,7.5) (15.6,7.5);
\node[anchor=west,font=\bfseries] at (.15,7.15) {A spatial benchmark for subcritical neutron counting};
\fill[gray!8] (1.2,3.0)--(4.3,3.0)--(5.3,3.8)--(5.3,6.2)--(2.2,6.2)--(1.2,5.4)--cycle;
\fill[blue!20] (1.2,3.0)--(2.2,3.8)--(2.2,6.2)--(1.2,5.4)--cycle;
\fill[orange!25] (4.3,3.0)--(5.3,3.8)--(5.3,6.2)--(4.3,5.4)--cycle;
\draw[thick] (1.2,3.0) rectangle (4.3,5.4);
\draw[thick] (1.2,5.4)--(2.2,6.2)--(5.3,6.2)--(5.3,3.8)--(4.3,3.0);
\draw[thick] (4.3,5.4)--(5.3,6.2);
\draw[dashed,gray] (1.2,3.0)--(2.2,3.8)--(5.3,3.8) (2.2,3.8)--(2.2,6.2);
\draw[<->] (1.2,2.65)--node[below] {$2L$: cube side} (4.3,2.65);
\node[anchor=east,text=blue!65!black,align=right] at (1.0,5.55) {$x=-L$\\left face};
\node[anchor=west,text=orange!65!black,align=left] at (5.4,5.55) {$x=+L$\\right face};
\fill[green!45!black] (2.65,4.15) circle (2pt);
\node[anchor=east,align=center,font=\fontsize{8}{9.5}\selectfont] at (2.5,4.05) {source\\event};
\draw[thick,green!40!black,->] (2.65,4.15)--(2.8,4.45)--(2.66,4.7)--(3.04,4.86)--(3.35,4.7);
\fill[red!75!black] (3.35,4.7) circle (2.3pt);
\node[above,font=\fontsize{8}{9.5}\selectfont] at (3.7,5.02) {fission};
\draw[thick,orange!80!black,->] (3.35,4.7)--(3.65,4.35)--(3.85,4.5)--(4.12,4.2)--(4.72,4.35);
\draw[thick,blue!65!black,->] (3.35,4.7)--(3.05,5.05)--(2.8,4.94)--(2.6,5.2)--(1.76,5.15);
\draw[thick,gray,->] (2.65,4.15)--(2.95,3.85)--(3.3,3.95);
\draw[red!65!black,thick] (3.3,3.85)--(3.5,4.05) (3.3,4.05)--(3.5,3.85);
\node[below,font=\fontsize{8}{9.5}\selectfont] at (3.5,3.85) {capture};
\node[align=center,font=\fontsize{8}{9.5}\selectfont] at (3.1,1.65) {All six faces absorb escaping particles.\\Only the two coloured faces are counted.};
\node[box,anchor=north west,text width=7.3cm] (inputs) at (7.0,6.65) {
\textbf{Inputs and physical meaning}\\[3pt]
$D$: diffusion coefficient (length$^2$/time).\\
$\lambda$: absorption/reproduction rate (1/time).\\
$q_j$: probability of $j$ descendants per event.\\
$\nu_1=\sum_j j q_j$: mean number of descendants.\\
$\rho=\lambda(\nu_1-1)$: net growth rate (1/time).\\
$S$: Poisson source events per unit time.\\
$p(\mathbf x)$: normalized source density, here\\
$p(\mathbf x)\propto\prod_{i=1}^3\cos(\pi x_i/(2L))$.\\
$\varepsilon$: probability of counting an exit on a coloured face.\\[3pt]
$a=D(\pi/2L)^2$: leakage scale (1/time).\\
$N_R(T),N_L(T)$: counts in a gate of duration $T$.};
\node[box,anchor=north west,text width=6.6cm,fill=blue!4] (sum) at ([xshift=-6.85cm,yshift=-.45cm]inputs.south west) {
\textbf{Sum: $N_R+N_L$} (slowest mode)\\
Symmetric decay: $\gamma_+=3a-\rho$.};
\node[box,anchor=north west,text width=6.6cm,fill=orange!5] (difference) at ([xshift=.7cm]sum.north east) {
\textbf{Difference: $N_R-N_L$} (slowest mode)\\
Antisymmetric decay: $\gamma_-=6a-\rho$.};
\node[box,anchor=north west,text width=14.4cm] at ([yshift=-.4cm]sum.south west) {
\textbf{Same fundamental decay, different leakage: an observable distinction}\\[2pt]
Reduced units: $L=1$, $\lambda=4$, $S=2$, $\varepsilon=1$. Both cases have $\gamma_+=0.6$.\\
Case A: $a=1$, $\gamma_-=3.6$. OpenMC gives $\widehat a=1.011\pm0.022$.\\
Case B: $a=1.3$, $\gamma_-=4.5$. OpenMC gives $\widehat a=1.308\pm0.028$.\\
Each $\pm$ is the standard error of the mean over 24 independent records, not a single-record uncertainty.};
\end{tikzpicture}}
\caption{The cube model. A source particle starts a chain: particles
diffuse, are captured or reproduce, and are lost when they reach a face. In
Secs.~\ref{sec:theory}--\ref{sec:openmc} every face is a detector. The
lower panels describe the opposite-face variant of
Sec.~\ref{sec:inversebenchmark}, in which all faces still absorb but only
the two coloured faces are observed. The rates $\gamma_\pm$ are those of the
diffusion model. In the corresponding OpenMC check the physical half-side is
$L-0.7104/\Sigma_t$, so that the extrapolated half-side is $L$, and fissions
are binary and analog, without FREYA.}
\label{fig:benchmark}
\end{figure}

\subsection{Counts in a gate: backward equation}

The measurement counts the particles recorded during a gate, the time
interval $[0,T]$. Since the chains started by different source particles are
independent, it suffices to characterize the counts produced by one
particle and its descendants. Following Ref.~\citep{zoia2014clustering}, we
do so by conditioning on the first event. Let $C_{s,\vx}$ be the number of
counts in $[0,T]$ among the descendants (itself included) of a particle
located at $\vx$ at time $s$, and let $W_s(z|\vx)=\E[z^{C_{s,\vx}}]$ be its
generating function.
During $[s,s+\dd s]$ the particle either diffuses or reproduces, and the
subtrees of its offspring are independent, so that
\begin{equation}
  -\partial_sW_s = D\nabla^2_{\vx}W_s + \lambda\left[G(W_s)-W_s\right],
  \qquad \vx\in V,\ s<T,
  \label{eq:backward}
\end{equation}
with the final condition $W_T=1$: a particle present at the end of the gate
produces no count. The detector enters through the boundary condition. A
particle on $\partial V$ at time $s$ has no descendants, and is recorded
with probability $\varepsilon$ if $s$ lies in the gate:
\begin{equation}
  W_s(z|\vx) = 1-\varepsilon+\varepsilon\,z^{\mathbf{1}_{[0,T]}(s)},
  \qquad \vx\in\partial V.
  \label{eq:bc}
\end{equation}
Equation~\eqref{eq:backward} is the equation of Appendix A of
Ref.~\citep{zoia2014clustering}. The only difference is the observable,
which now sits in a time-dependent boundary condition instead of the
initial condition.

Only the first two moments are needed. Differentiating once and twice with
respect to $z$ at $z=1$, the mean $f_s(\vx)=\partial_zW_s|_{z=1}$ and the
second factorial moment $g_s(\vx)=\partial^2_zW_s|_{z=1}$ obey the linear
equations
\begin{align}
  -\partial_s f_s &= \Lop f_s, &
  f_s|_{\partial V} &= \varepsilon\,\mathbf{1}_{[0,T]}(s), & f_T&=0,
  \label{eq:f}\\
  -\partial_s g_s &= \Lop g_s + \lambda\nu_2\,f_s^2, &
  g_s|_{\partial V} &= 0, & g_T&=0,
  \label{eq:g}
\end{align}
with the backward operator $\Lop=D\nabla^2+\rho$. Let
$\mathcal{G}_\theta(\vx',\vx)$ be the Green's function of
$\partial_\theta-\Lop$ with homogeneous Dirichlet conditions. Two mean
quantities follow from it: the mean number of descendants present in $V$ a
time $\theta$ after the birth of their ancestor at $\vx$, and their mean
leakage rate,
\begin{equation}
  m_\theta(\vx) = \int_V\mathcal{G}_\theta(\vx',\vx)\,\dd\vx',\qquad
  J_\theta(\vx) = \rho\,m_\theta(\vx) - \partial_\theta m_\theta(\vx).
  \label{eq:J}
\end{equation}
The second relation is the particle balance: what is produced and does not
remain in $V$ has leaked. By Duhamel's principle, as in Eq.~(A10) of
Ref.~\citep{zoia2014clustering},
\begin{align}
  f_s(\vx) &= \varepsilon\int_{\max(s,0)}^{T} J_{s'-s}(\vx)\,\dd s', \label{eq:fsol}\\
  g_s(\vx) &= \lambda\nu_2\int_s^T\dd s'\int_V\dd\vx'\,
  \mathcal{G}_{s'-s}(\vx',\vx)\,f_{s'}(\vx')^2 . \label{eq:gsol}
\end{align}
Equation~\eqref{eq:fsol} integrates the leakage rate over the part of the
gate that follows $s$. Equation~\eqref{eq:gsol} has the familiar
interpretation of Ref.~\citep{zoia2014clustering}: two recorded particles
have a last common ancestor, which reproduced at $(\vx',s')$, and the two
branches then reached the detector independently.

\subsection{Stationary Poisson source}

Source particles are emitted one at a time by a stationary Poisson process
of rate $S$, at independent positions of density $p(\vx)$ normalized on $V$.
We now denote by $C$ the total number of counts in the gate. The source has
been operating since $s=-\infty$ and the system is subcritical: the largest
eigenvalue of $\Lop$ is strictly negative. The counts of the gate then form
a Poisson cluster process, and Campbell's theorem expresses their mean and
variance as sums over source events,
\begin{equation}
  \E[C]=S\int_{-\infty}^{T}\!\dd s\int_V p\,f_s,\qquad
  \Var[C]=S\int_{-\infty}^{T}\!\dd s\int_V p\,(f_s+g_s).
\end{equation}
The Feynman $Y$ function is the excess of the variance-to-mean ratio over
its Poisson value. It vanishes for uncorrelated counts, and it does not
depend on the source intensity:
\begin{equation}
  Y(T)\equiv\frac{\Var[C]}{\E[C]}-1
  = \frac{\int\dd s\int p\,g_s}{\int\dd s\int p\,f_s}.
\end{equation}
The denominator is $\varepsilon T\langle h\rangle$, where
$h(\vx)=\int_0^\infty J_\theta(\vx)\,\dd\theta$ is the mean number of leaks of a
chain started at $\vx$ and $\langle h\rangle=\int p\,h$. In the numerator, the
integrals over the source position and over $s$ act on the Green's function
only. They produce the stationary particle density per unit source rate,
\begin{equation}
  \widehat c(\vx') = \int_0^\infty\dd\theta\int_V\dd\vx\,p(\vx)\,
  \mathcal{G}_\theta(\vx',\vx),\qquad
  D\nabla^2\widehat c+\rho\,\widehat c+p=0,\quad \widehat c|_{\partial V}=0 ,
\end{equation}
so that $S\widehat c$ is the stationary particle concentration. We obtain
the general result
\begin{equation}
  \boxed{\;
  Y(T) = \frac{\lambda\nu_2}{\varepsilon\,T\,\langle h\rangle}
  \int_V\dd\vx\,\widehat c(\vx)\int_{-\infty}^{T}f_s(\vx)^2\,\dd s\;}
  \label{eq:Ygeneral}
\end{equation}
valid for any bounded domain and any normalized source density under the
assumptions above. The integrand reads directly: $\lambda\nu_2\widehat c$ is
the local rate of pair production per unit source intensity, and $f_s^2$ is
the product of the mean counts produced in the gate by the two independent
branches of a pair born at time $s$.

\paragraph{Long gates.}
For long gates $f_s\to\varepsilon h$ over most of the gate, and $Y$ reaches
the plateau
\begin{equation}
  Y_\infty = \varepsilon\,\lambda\nu_2\,
  \frac{\int_V\widehat c\,h^2}{\int_V p\,h}.
  \label{eq:Yinf}
\end{equation}
The square $h^2$ is essential. A fission occurring where both descendants
are likely to reach the boundary contributes more than one occurring where
they are likely to be lost, so the concentration alone cannot determine the
counting variance. The limits $\nu_2=0$ and $\varepsilon\to0$ give $Y=0$,
as expected for independent particles and for rare detection.

Point kinetics is recovered by ignoring space. With a uniform
$\widehat c=1/\gamma$ and $h=b/\gamma$, for a leakage rate $b$ and a decay
constant $\gamma=b-\rho$, Eq.~\eqref{eq:Yinf} reduces to the prompt
Feynman-$\alpha$ asymptote $Y_\infty=\varepsilon\,b\,\lambda\nu_2/\gamma^2$.
This is the familiar $\lambda_d\lambda_f\langle\nu(\nu-1)\rangle/\alpha^2$,
with fission rate $\lambda_f$, detection rate $\lambda_d=\varepsilon b$ and
$\alpha=-\gamma$ per neutron~\citep{pazsit2007neutron}.

The plateau can also be computed without any time dependence. Let $h$ and
$w$ solve
\begin{equation}
 \Lop h=0,\quad h|_{\partial V}=1,\qquad
 \Lop w=-h^2,\quad w|_{\partial V}=0.
 \label{eq:stationary}
\end{equation}
Since $-\Lop\widehat c=p$, Green's identity gives
$\int_V\widehat c h^2=\int_Vp w$, and hence
$Y_\infty=\varepsilon\lambda\nu_2\int_Vp w/\int_Vp h$. Only two linear
boundary-value problems are required. Once $h$ and $w$ are known, different
source distributions can be compared by a simple integration, without
repeating a branching simulation.

\subsection{Cube}

In the cube $V=[-L,L]^3$ with absorbing walls, the eigenfunctions of $\Lop$
are products of the one-dimensional Dirichlet modes
\begin{equation}
  \psi_n(x)=L^{-1/2}\sin\frac{n\pi(x+L)}{2L},\qquad
  \varphi_{\vn}(\vx)=\prod_{i=1}^3\psi_{n_i}(x_i),\qquad
  \alpha_{\vn}=\rho-a|\vn|^2,\quad a=D\left(\frac{\pi}{2L}\right)^2,
\end{equation}
with $n_i\ge1$, and
$\mathcal{G}_\theta(\vx,\vx')=\sum_{\vn}\varphi_{\vn}(\vx)\varphi_{\vn}(\vx')e^{\alpha_{\vn}\theta}$.
Each mode $\vn$ has a leakage rate $b_{\vn}=a|\vn|^2$ and a decay constant
$\gamma_{\vn}=-\alpha_{\vn}=b_{\vn}-\rho>0$. The whole spectrum is thus fixed
by two numbers, the leakage scale $a$ and the net growth rate $\rho$. The
fundamental mode $\vn=(1,1,1)$ has $b_1=3a$ and $\gamma_1=3a-\rho$. The
one-group multiplication factor is the ratio of production to losses in this
mode, $k=\lambda\nu_1/(\lambda+b_1)$.

When the whole boundary is a detector, only modes that are odd in each
index have a nonzero volume integral. For odd $n$, define
\begin{equation}
  c_n=\int_{-L}^{L}\psi_n=\frac{4\sqrt L}{n\pi},\qquad
  \tau_{nkl}=\int_{-L}^{L}\psi_n\psi_k\psi_l
  =\frac{1}{2\sqrt L}\left[s_{n+k-l}+s_{n-k+l}+s_{-n+k+l}-s_{n+k+l}\right],
\end{equation}
with $c_n=0$ for even $n$ and
$s_j=(1-\cos j\pi)/(j\pi)$ ($s_0=0$). We write $C_{\vn}=\prod_ic_{n_i}$ for
the volume integral of mode $\vn$,
$T_{\vn\bm k\bm l}=\prod_i\tau_{n_ik_il_i}$ for the overlap of three modes,
and $S_{\vn}=\int p\,\varphi_{\vn}$ for the projection of the source
($S_{\vn}=C_{\vn}/|V|$ for a uniform source). Equation~\eqref{eq:J} gives
$J_\theta=\sum_{\vn}C_{\vn}b_{\vn}\varphi_{\vn}e^{-\gamma_{\vn}\theta}$, hence
$\langle h\rangle=\sum_{\vn}S_{\vn}C_{\vn}b_{\vn}/\gamma_{\vn}$. Inserting
Eq.~\eqref{eq:fsol} into Eq.~\eqref{eq:Ygeneral},
\begin{equation}
  Y(T)=\frac{\varepsilon\lambda\nu_2}{T\langle h\rangle}
  \sum_{\vn,\bm k,\bm l}\frac{S_{\vn}}{\gamma_{\vn}}\,
  C_{\bm k}b_{\bm k}\,C_{\bm l}b_{\bm l}\,T_{\vn\bm k\bm l}\,
  F(\gamma_{\bm k},\gamma_{\bm l};T).
  \label{eq:Ycube}
\end{equation}
In this sum, mode $\vn$ carries the stationary density that produces the
pairs, and modes $\bm k$ and $\bm l$ carry the two branches to the boundary.
The function $F$ is the time integral of the two gate responses. Writing
$H(y)=1-(1-e^{-y})/y$,
\begin{equation}
 F(\gamma_k,\gamma_l;T)=\frac{T}{\gamma_k\gamma_l}
 \frac{\gamma_l H(\gamma_kT)+\gamma_k H(\gamma_lT)}{\gamma_k+\gamma_l}.
 \label{eq:F}
\end{equation}
It includes ancestors born before the gate as well as those born within it.
For $k=l$, $F=T H(\gamma_kT)/\gamma_k^2$, the usual Feynman shape. More
generally, Eq.~\eqref{eq:Ycube} is a sum of Feynman functions with different
decay constants, as in the multimode descriptions of
Refs.~\citep{munoz2011,degweker2016}. Here the coefficients are specified by
the source and the absorbing boundary, and are not fitted independently.

\paragraph{Fundamental mode.}
Keeping only $\vn=\bm k=\bm l=(1,1,1)$ in Eq.~\eqref{eq:Ycube} and in
$\langle h\rangle$ gives
\begin{equation}
  Y_{\mathrm{f}}(T)=\kappa\,\frac{\varepsilon\,b_1\lambda\nu_2}{\gamma_1^2}
  \left[1-\frac{1-e^{-\gamma_1T}}{\gamma_1T}\right],\qquad
  \kappa=c_1^3\,\tau_{111}^3=\left(\frac{32}{3\pi^2}\right)^3\simeq1.262,
  \label{eq:Yfund}
\end{equation}
whatever the source, provided its projection on the fundamental mode is
nonzero. This is the point-kinetics formula multiplied by the form factor
$\kappa=\int\varphi_1\int\varphi_1^3$ of the normalized fundamental mode
$\varphi_1\equiv\varphi_{(1,1,1)}$. The factor expresses the spatial overlap between the fundamental density
mode and the two descendant responses, relative to the mean count response. It equals one for a flat mode.

\paragraph{Higher modes.}
The fundamental mode is not the whole story. For a uniform source it
carries a fraction $C_1^2/|V|=(8/\pi^2)^3\simeq0.53$ of the initial mean
population, with $C_1\equiv C_{(1,1,1)}$. This is a spectral weight, not a
partition of particles into separate populations. The other modes describe,
in particular, the rapid escape of particles born near the walls. Their
weight in the mean count differs from their weight in the pairs. They can
therefore change the plateau of $Y(T)$ as well as its shape, and the sign of
the correction is not fixed in general. In the critical limit
$\gamma_1\to0^+$, at fixed geometry and nonzero source projection, the
fundamental term dominates: the ratio of the long-gate BBM prediction to the
uncorrected point-model prediction tends to $\kappa$, not to one.

\paragraph{Numerical evaluation.}
For a separable source, the matrix
$A_{\bm k\bm l}=\int_V\widehat c\,\varphi_{\bm k}\varphi_{\bm l}$
can be evaluated through one-dimensional heat kernels and a Laplace
integral (Supplementary Material, Sec.~S2), which avoids a direct triple sum
over three-dimensional modes. The sum is truncated to odd indices
$n_i\le2m-1$. The curves shown below use $m=11$ and $m=15$, extrapolated
linearly in $1/m$.
The fits of Sec.~\ref{sec:k} use $m=7$. Repeating them with $m=5$ and $9$
reveals a non-negligible dependence on the truncation, quantified in
Supplementary Material, Sec.~S3. The truncation error is a numerical error
of the theory and must be kept distinct from the Monte Carlo uncertainty of
the simulations.

\section{Check against direct BBM simulations}
\label{sec:bbm-mc}

We first check Eq.~\eqref{eq:Ycube} against a direct simulation of the same
process, so that any difference can only come from the numerical evaluation
of either side.

\paragraph{Set-up.}
The parameters are $L=D=1$ and $\lambda=10$, with the FREYA multiplicity of
thermal $^{235}$U fission. The fraction of absorptions that are fissions is
adjusted to obtain $k=0.5$, $0.8$ and $0.95$, which in the real system would correspond to varying the enrichment. Each of the $8\times10^5$
chains starts from one particle placed uniformly in the cube. Particles
move by Gaussian increments over a time step $\dd t=10^{-4}$. A crossing of
a face during a step is detected with the Brownian-bridge probability
$\exp[-d_0d_1/(D\,\dd t)]$, where $d_0$ and $d_1$ are the distances to the
face at the beginning and at the end of the step. The chains are assigned
independent uniform start times in a finite recording interval, and leakage
times are binned sequentially. Error bars are estimated from $32$ segments
of the record.

Two features of this simulation differ from the theory. A fixed number of
uniformly timed chains is a Poisson source conditioned on its total number
of events, not an unconditioned stationary process (Supplementary Material,
Sec.~S3). And the time step is finite: the bridge correction does not by
itself make the sampled leakage times exact.

\begin{figure}[t]
\centering
\includegraphics[width=\textwidth]{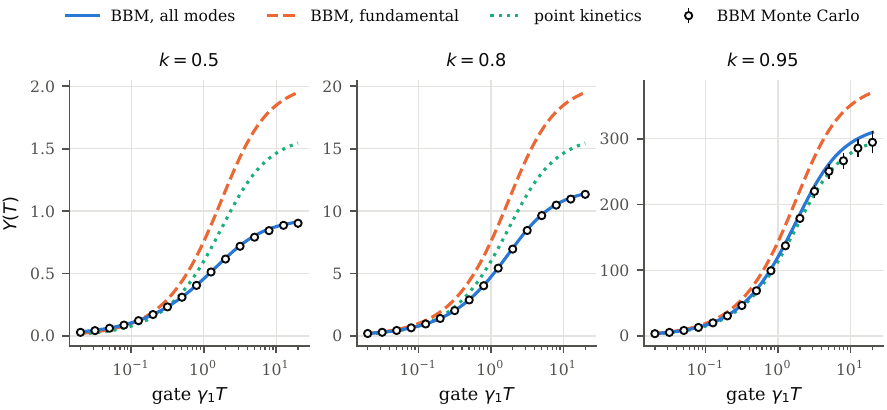}
\caption{Feynman $Y$ of the leakage counts of a branching Brownian motion in
a cube: direct simulation with uniformly timed source histories, Eq.~\eqref{eq:Ycube} with all modes, the fundamental
mode Eq.~\eqref{eq:Yfund}, and point kinetics with the same parameters.}
\label{fig:bbm}
\end{figure}

\paragraph{Results.}
Figure~\ref{fig:bbm} shows that Eq.~\eqref{eq:Ycube} reproduces the simulated
$Y(T)$ over three decades of gates. The mean number of leaks per source
particle is $0.909$, $1.639$ and $5.02\pm0.04$ in the simulation, against
$0.910$, $1.639$ and $5.07$ in theory. For $Y(T)$ the agreement depends on
$k$:
\begin{itemize}
\item at $k=0.8$, within $1\%$ for all displayed gates;
\item at $k=0.5$, within $2\%$ for $\gamma_1T\ge0.05$; the two shortest
  gates, which span only about twenty time steps, deviate by $7$ and $12\%$;
\item at $k=0.95$, the simulation lies $2$ to $5\%$ below the theory, that
  is, $1.5$ to $2$ standard deviations. The offset is strongly correlated
  between gates, since all of them are computed from the same record.
\end{itemize}
These are pointwise discrepancies, not a covariance-based goodness-of-fit
test. Runs at $\dd t=2\times10^{-4}$ and $5\times10^{-5}$ are compared in
Supplementary Material, Sec.~S3. Their scatter does not establish
time-step convergence, particularly near criticality, so the residuals at
$k=0.95$ cannot yet be attributed to statistics alone.

\paragraph{Simplified models.}
Neither simplified model is reliable. The fundamental mode,
Eq.~\eqref{eq:Yfund}, overestimates the long-gate plateau by factors of
$2.15$, $1.71$ and $1.20$ at $k=0.5$, $0.8$ and $0.95$. Point kinetics with
the exact one-group parameters ($\kappa=1$) overestimates it by factors of
$1.70$ and $1.36$ at $k=0.5$ and $0.8$, and underestimates it by $5\%$
(factor $0.95$) at $k=0.95$.
The near agreement of the point model at $k=0.95$ is a compensation at these
particular parameters: its error has changed sign between $k=0.8$ and $0.95$, on the way to the
critical limit, where it underestimates the plateau by the factor
$1/\kappa\simeq0.79$. A separate comparison of population moments is given
in Supplementary Material, Sec.~S4.

\section{Analog fission in OpenMC}
\label{sec:analog}

To compare the model with transport, the Monte Carlo code must reproduce
not only the mean fission production but also the number of neutron pairs
created at each fission, since $Y$ is proportional to the pair-production
rate $\lambda\nu_2$. A simulation with this property, in which every
history is a physically possible one and carries unit weight, is called
\emph{analog}. The default fission treatment of OpenMC is not analog in
this sense. We first quantify the consequence, then describe the analog
option that we added.

\subsection{Why the standard algorithm is not analog}

In the expected-value collision treatment considered here,
OpenMC~\citep{romano2015openmc,openmcphysics} banks fission sites before
sampling absorption. Without survival biasing or other weight adjustments,
every collision with a fissionable nuclide banks a number of fission sites
\begin{equation}
  n = \lfloor x\rfloor + \mathbf{1}\{\xi < x - \lfloor x\rfloor\},
  \qquad x = \frac{w}{k}\,\frac{\nub\,\sigma_f}{\sigma_t},
  \label{eq:openmc-nu}
\end{equation}
where $\xi$ is a uniform random number, $w$ the particle weight and $k$ the
current multiplication factor, equal to one in fixed-source mode. Then,
independently, the particle is absorbed with probability
$\sigma_a/\sigma_t$; otherwise it scatters. This algorithm preserves the
expected fission production, but it alters the fluctuations in two ways.
\begin{enumerate}
\item The multiplicity is restricted to the two integers that bracket $x$.
  When $w=k=1$ and $\nub\sigma_f/\sigma_t<1$, a collision creates at most
  one fission neutron.
\item Fission neutrons are created at collisions that may turn out to be
  scatterings, and the incident neutron may survive a ``fission''. In the
  restricted case above, the only pairs created at a collision consist of a
  surviving parent and its single child.
\end{enumerate}

Both effects can be quantified for a homogeneous one-speed collision law
with unit weights, $k=1$ and $x<1$. The survival and banking draws are
independent under this law; in a mixture the calculation would have to be
conditioned on the sampled nuclide. The offspring of a collision consist of
the surviving parent ($s=1$ with probability $\Sigma_s/\Sigma_t$) and of the
banked sites ($n$, Bernoulli with parameter $x=\nub\Sigma_f/\Sigma_t$). At
the collision rate $v\Sigma_t$, the first two factorial moments give a net
growth rate and a pair-production rate
\begin{align}
  v\Sigma_t\,\E[s+n-1] &= v(\nub\Sigma_f - \Sigma_a), \\
  v\Sigma_t\,\E[(s+n)(s+n-1)] &= 2\,v\,\frac{\Sigma_s\,\nub\Sigma_f}{\Sigma_t}.
  \label{eq:std-pairs}
\end{align}
An analog fission, occurring at rate $v\Sigma_f$ and emitting $\nu$
neutrons, gives the same net growth rate but a pair-production rate
$v\Sigma_f\langle\nu(\nu-1)\rangle$. In the language of
Sec.~\ref{sec:theory}, the standard algorithm is a branching process in
which collisions are the reproduction events: it has the correct $\rho$ and
a different $\lambda\nu_2$. Since $Y$ is proportional to $\lambda\nu_2$
[Eq.~\eqref{eq:Ygeneral}], the ratio of the two pair-production rates,
\begin{equation}
  R = \frac{2\,\Sigma_s/\Sigma_t}{\nub\,D_2},\qquad
  D_2=\frac{\langle\nu(\nu-1)\rangle}{\nub^2},
  \label{eq:ratio}
\end{equation}
is also the ratio of the Feynman $Y$ functions of the two diffusion models,
provided their mean operator, source and detector are identical. For
thermal $^{235}$U fission, with $D_2\simeq0.79$, $\nub\simeq2.42$ and
$\Sigma_s/\Sigma_t=0.9$, this gives $R\simeq0.94$: the standard algorithm
underestimates $Y$ by $6\%$.

This deficit is not universal. $R$ can exceed one, and a single scalar
correction does not in general account for energy and angular correlations.
For arbitrary $x$, writing $r=x-\lfloor x\rfloor$, the banked sites add
$\E[n(n-1)]=x^2-x+r(1-r)$ to the collision pair moment.
Equation~\eqref{eq:ratio} is therefore used only for the one-speed
comparisons with $x<1$.

\subsection{Analog collision algorithm}

The analog mode, enabled by the new setting \texttt{analog\_fission},
replaces the treatment of the absorption channel. A single random number
selects fission, capture or scattering with probabilities
$\sigma_f/\sigma_t$, $(\sigma_a-\sigma_f)/\sigma_t$ and
$\sigma_s/\sigma_t$. A fission kills the incident neutron and banks $\nu$
unit-weight neutrons at the collision point and time. The multiplicity
$\nu$ is drawn from the model selected by \texttt{analog\_fission\_model}:
\begin{description}
\item[\texttt{bernoulli}] the rounded multiplicity of
  Eq.~\eqref{eq:openmc-nu} with $x=\nub$, which isolates the effect of the
  multiplicity distribution;
\item[\texttt{llnl}] the tabulated multiplicity distribution of the LLNL
  Fission library;
\item[\texttt{freya}] a complete FREYA fission event.
\end{description}
Their definitions, the implementation and the sampled multiplicity moments
are given in Supplementary Material, Sec.~S1.

The implementation uses the LLNL Fission library v2.0.5, restricted to the
uranium reactions~\citep{llnllibrary}. The multiplicity is read from the
same event as its outgoing neutrons. Calls are serialized, because the
library stores the event in global state. Three restrictions apply.
\begin{itemize}
\item \emph{Fixed source only.} The analog option is intended for
  fixed-source histories. Population control in an eigenvalue calculation
  would not preserve physical counting fluctuations. Survival biasing and
  weight windows are incompatible with the analog mode and are rejected at
  initialization.
\item \emph{Prompt neutrons only.} All simulations below are prompt only.
  The optional independent Poisson treatment of delayed neutrons is an
  approximation, not a verification of prompt--delayed correlations.
\item \emph{Multigroup calculations.} Only the multiplicity is taken from
  FREYA: outgoing groups follow the material spectrum and directions are
  isotropic. These calculations therefore test the multiplicity treatment,
  not FREYA's energy--angle correlations.
\end{itemize}
One point matters for the comparisons below. FREYA computes its own mean
multiplicity: for $^{238}$U fission at $2$~MeV it is about $1.9\%$ below the
evaluated mean. Simulations that use different fission models therefore
need not have the same mean multiplication.

\section{Comparison with OpenMC transport}
\label{sec:openmc}

We now compare the BBM prediction with transport simulations of the same
cube: first one-speed transport, where diffusion is the principal physical approximation
(Sec.~\ref{sec:onespeed}), then the inference of $k$ from the Feynman curve
(Sec.~\ref{sec:k}), and finally continuous-energy transport, where the
one-group model reaches its limit (Sec.~\ref{sec:ce}).

In all simulations the cube has vacuum boundaries and every leaking neutron
is recorded ($\varepsilon=1$), by writing the surface source file of the six
faces. The source is uniform in the cube, with emission times uniform over a
measurement window $[0,T_m]$. $T_m$ is chosen so that about two neutrons are
counted per decay time $1/\gamma_1$, and the build-up phase $t<20/\gamma_1$
is discarded. As in Sec.~\ref{sec:bbm-mc}, fixing the number of uniformly
timed histories conditions the source count; the prediction is the
stationary-source formula and does not include finite-record effects. The
leakage times are binned sequentially for $30$ gate durations. The
uncertainties are batch means over $8$ segments of each of $4$ independent
replicas. Delayed neutrons are disabled.

\subsection{One-speed cube}
\label{sec:onespeed}

To test the diffusion description against one-speed transport, we use OpenMC
in multigroup mode with a single group: $\Sigma_t=1$~cm$^{-1}$,
$\Sigma_a=0.1$~cm$^{-1}$, isotropic scattering, $v=10^9$~cm\,s$^{-1}$ and
the multiplicity of thermal $^{235}$U fission. The cube has a physical
half-side of $15$~cm, so its side is $30$ mean free paths.
The BBM parameters follow from Eq.~\eqref{eq:mapping} without any
adjustment. The Dirichlet condition is applied on the extrapolated boundary,
so that the half-side of the diffusion model is
$L=15~\mathrm{cm}+0.7104/\Sigma_t=15.71$~cm.
The source stays uniform in the physical cube, and $S_{\vn}$ is the
corresponding projection. $\nub\Sigma_f$ is chosen to give a diffusion
multiplication factor $k=0.5$, $0.8$ and $0.9$. OpenMC eigenvalue
calculations give $k=0.5005$, $0.8008$ and $0.9004$, and pulsed-source
simulations give decay constants within $1$--$6\%$ of $\gamma_1$. The
diffusion model thus reproduces the fundamental decay rate to this accuracy.

\begin{figure}[t]
\centering
\includegraphics[width=\textwidth]{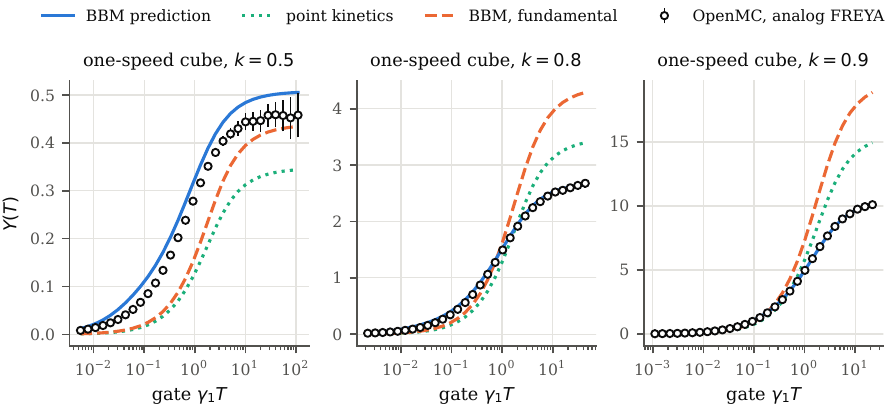}
\caption{One-speed cube simulated with OpenMC (analog fission, FREYA
multiplicity of thermal $^{235}$U), versus the parameter-free BBM prediction,
its fundamental mode and point kinetics. Gates are scaled by the
fundamental decay constant $\gamma_1$ of the diffusion model.}
\label{fig:mg}
\end{figure}

\paragraph{Feynman curves.}
Figure~\ref{fig:mg} compares the analog FREYA simulations with the
prediction of Eq.~\eqref{eq:Ycube}, which has no free parameter.
Table~\ref{tab:plateau} summarizes the deviations at the longest simulated
gate, $T=2~\mu$s. This finite-gate comparison does not assume that the
asymptotic plateau has already been reached exactly.

\begin{table}[t]
\centering
\caption{One-speed cube: signed relative deviations $100(Y_{\rm model}/Y_{\rm MC}-1)$
at the longest simulated gate, $T=2~\mu$s, with the same parameters.
The second row gives the maximum absolute relative deviation over all
stored gates with $T>10/(v\Sigma_t)$. The values use the stored simulation
means and extrapolated modal predictions.}
\label{tab:plateau}
\begin{tabular}{lccc}
\toprule
 & $k=0.5$ & $k=0.8$ & $k=0.9$ \\
\midrule
BBM, all modes ($T=2~\mu$s) & $+10.3\%$ & $-1.0\%$ & $+0.24\%$ \\
BBM, all modes (largest, $T>10/v\Sigma_t$) & $15.9\%$ & $6.9\%$ & $3.6\%$ \\
BBM, fundamental mode & $-5\%$ & $+60\%$ & $+87\%$ \\
Point kinetics & $-25\%$ & $+27\%$ & $+48\%$ \\
\bottomrule
\end{tabular}
\end{table}

The BBM deviates from the simulation at gates of a few collision times,
where the neutrons have not yet reached the diffusive regime. The larger
discrepancies at low $k$ are consistent with an increased contribution of
short chains near the vacuum boundary: diffusion cannot resolve this
transport layer, even when the low-order decay constants agree. The
percentages are observed discrepancies in these configurations, not error
bounds for arbitrary sources or geometries.

\begin{figure}[t]
\centering
\includegraphics[width=0.5\textwidth]{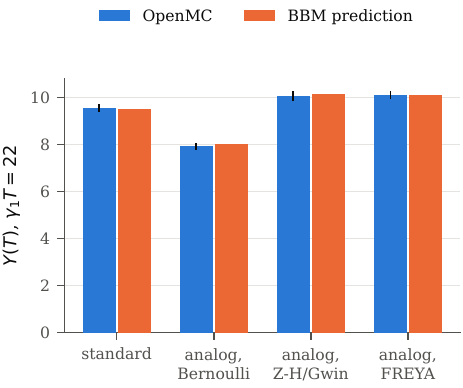}
\caption{Long-gate $Y$ of the one-speed cube ($k=0.9$) for the four fission
algorithms, and the corresponding BBM predictions.}
\label{fig:modes}
\end{figure}

\paragraph{Fission algorithms.}
Figure~\ref{fig:modes} compares the four fission algorithms at $k=0.9$:
standard OpenMC and the three analog models. The comparison isolates the
pair-production rate, given by Eq.~\eqref{eq:mapping} with the sampled
moments in the analog modes and by Eq.~\eqref{eq:std-pairs} in the standard
mode. At the longest gate, the BBM reproduces all four values within $1.2\%$.
For standard OpenMC, Eq.~\eqref{eq:ratio} predicts a deficit of about $6\%$
relative to analog FREYA, compared with $5.3\%$ in the simulations
($9.56$ simulated, $9.52$ predicted for standard OpenMC). Analog Bernoulli
rounding gives a deficit of about $21\%$ ($7.94$ simulated, $8.03$ predicted).
The tabulated and FREYA distributions give the same result within the
displayed uncertainty. Their sampled moments are close but not identical
(Supplementary Material, Table~S1); a precise comparison must also retain
any difference in the mean production rate.

\subsection{From the Feynman curve to $k$}
\label{sec:k}

So far the material and the source were prescribed and $Y(T)$ was
predicted. We now turn the question around: if these quantities are known
but the fundamental decay rate is not, does retaining the spatial modes
improve its estimate from a measured $Y(T)$?

\paragraph{What is assumed known.}
The cube, the diffusion coefficient, the spatial source and the ideal
leakage detector are prescribed. They determine $a=D(\pi/2L)^2$ and all
source projections. Converting a decay rate into a multiplication factor
also requires the prompt generation time $\Lambda=1/(\lambda\nu_1)$. The
efficiency and the pair-production rate enter only through an overall
amplitude. This is therefore a conditional inverse problem. It does not
determine an unknown geometry, source distribution or generation time from a
single counting curve.

\paragraph{A two-parameter model.}
Once $a$ is known, the decay constants of all the modes follow from that of
the fundamental mode,
\begin{equation}
  \gamma_{\vn} = a|\vn|^2-\rho = \gamma_1 + a\left(|\vn|^2-3\right),
  \label{eq:harmonics}
\end{equation}
because all modes share the same net growth rate $\rho$. Through
$\rho=b_1-\gamma_1$, $\gamma_1$ also fixes $\langle h\rangle$ and the
overlap matrix $A_{\bm k\bm l}$. The whole shape of $Y(T)$ in
Eq.~\eqref{eq:Ycube} is therefore set by the single parameter $\gamma_1$,
and its amplitude by $\varepsilon\lambda\nu_2$. The BBM thus has the same
number of free parameters as the point-kinetics formula
$Y_\infty H(\gamma_{\mathrm{pt}}T)$. The difference lies in the shape:
instead of a single exponential, it contains the harmonics, with decay
constants locked to $\gamma_1$ by Eq.~\eqref{eq:harmonics}.

\paragraph{From $\gamma_1$ to $k$.}
In the one-group model, $k=\lambda\nu_1/(\lambda+b_1)$ is the ratio of the
production rate to the loss rate (absorption plus leakage) in the
fundamental mode. Hence
$\gamma_1=b_1+\lambda-\lambda\nu_1=\lambda\nu_1(1-k)/k$, that is,
\begin{equation}
  k=\frac{1}{1+\gamma_1\Lambda},
  \qquad\text{and}\qquad
  \frac{\delta k}{k}=-(1-k)\left(\frac{\delta\gamma_1}{\gamma_1}+\frac{\delta\Lambda}{\Lambda}\right).
  \label{eq:k-from-gamma}
\end{equation}
The second relation propagates small errors in the decay rate and in the
generation time. An error in $a$ or in the source also changes the fitted
$\gamma_1$.

\paragraph{Example.}
For the cube with reference $k=0.9004$, the extrapolated half-side of
$15.71$~cm gives $a=3.33\times10^6$~s$^{-1}$ and
$b_1=1.00\times10^7$~s$^{-1}$. The first contributing harmonic, $(1,1,3)$,
decays $8a$ faster than the fundamental. Fitting the $16$ gates from
$10$~ns to $2~\mu$s with $m=7$ gives $\gamma_1=1.030\times10^7$~s$^{-1}$,
against the diffusion value $1.100\times10^7$~s$^{-1}$. With
$\Lambda=1.010\times10^{-8}$~s, Eq.~\eqref{eq:k-from-gamma} gives
$k=0.9058$. The point-model fit gives
$\gamma_{\mathrm{pt}}=2.39\times10^7$~s$^{-1}$ and $k=0.805$: its single
time scale absorbs a contribution from the faster modes.
Table~\ref{tab:k} gives the estimates for the three cubes and for two
choices of the shortest gate. Retaining the spatial modes reduces the error
on $k$ from approximately $0.03$--$0.22$ to below $0.014$ in these
$m=7$ fits. These are absolute errors in $k$, not relative percentages.

\begin{table}[t]
\centering
\caption{Multiplication factors of the one-speed cube estimated from $Y(T)$
(analog FREYA runs) with Eq.~\eqref{eq:k-from-gamma}, for two choices of the
shortest gate used in the fit, with $m=7$. All inputs other than the decay
rate and the amplitude are prescribed. These are point estimates, not
confidence intervals.}
\label{tab:k}
\begin{tabular}{lccc}
\toprule
reference $k$ (OpenMC eigenvalue) & 0.5005 & 0.8008 & 0.9004 \\
\midrule
\multicolumn{4}{l}{\emph{gates $T>10/v\Sigma_t$}} \\
\quad BBM fit & 0.511 & 0.814 & 0.906 \\
\quad point-kinetics fit & 0.280 & 0.613 & 0.805 \\
\multicolumn{4}{l}{\emph{gates $T>100/v\Sigma_t$}} \\
\quad BBM fit & 0.502 & 0.811 & 0.901 \\
\quad point-kinetics fit & 0.322 & 0.717 & 0.867 \\
\bottomrule
\end{tabular}
\end{table}

\paragraph{How robust are these estimates?}
The nominal diagonal least-squares error would give
$\sigma_k\simeq2.5\times10^{-4}$ with all other inputs fixed. This is not a
measurement uncertainty, for two reasons: the gates share the same count
record, and the residuals include transport and truncation errors. A
reanalysis of the stored block estimates shows the size of these effects.
\begin{itemize}
\item \emph{Truncation.} Changing $m$ from $5$ to $7$ to $9$ gives
  $k=0.9047$, $0.9058$, $0.9056$ for the reference $0.9004$, and $0.4726$,
  $0.5110$, $0.5225$ for the reference $0.5005$. The particularly good
  agreement at $m=7$ is therefore partly a truncation effect.
\item \emph{Weighting.} Using the empirical inter-gate covariance instead of
  diagonal weights gives $0.9084$ at $m=9$. The covariance estimates are
  themselves noisy ($32$ blocks from four records); they serve as a
  sensitivity check, not as a calibrated goodness-of-fit test.
\end{itemize}
The estimates of Table~\ref{tab:k} thus show a substantial reduction of the
point-model bias, but neither gate window establishes a converged
inversion. The full sensitivity study is given in Supplementary Material,
Sec.~S3.

\subsection{Continuous-energy uranium cube}
\label{sec:ce}

The last configuration tests the one-group reduction itself. It is a
cube of uranium metal with a side of $50$~cm ($19.05$~g\,cm$^{-3}$) with $^{235}$U
enrichments of $3$, $7$ and $10.5\%$, simulated in continuous energy.
The eigenvalues computed with the evaluated data are
$k_{\mathrm{eval}}=0.513$, $0.748$ and $0.906$ (OpenMC, ENDF/B-VII.1, prompt
only). They are not eigenvalue references for the FREYA event model, which
can change both the multiplicity and the spectrum. The one-group BBM
parameters are taken from two sources: the late-time rates per neutron of
pulsed-source simulations (fission, absorption and leakage rates, normalized
by the neutron population obtained from the inverse-velocity tally), and the
analog multiplicity histograms.

\begin{figure}[t]
\centering
\includegraphics[width=\textwidth]{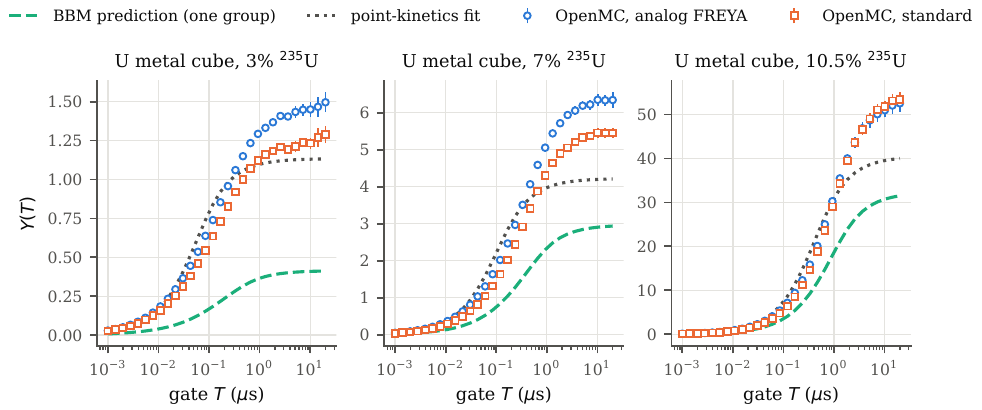}
\caption{Uranium-metal cube, continuous energy: OpenMC with analog FREYA
fission and with the standard algorithm, the one-group BBM prediction and a
point-kinetics fit.}
\label{fig:ce}
\end{figure}

\begin{figure}[t]
\centering
\includegraphics[width=0.5\textwidth]{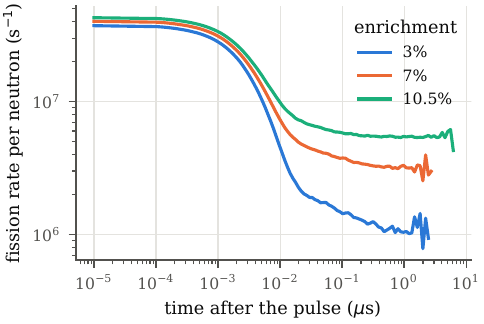}
\caption{Fission rate per neutron after a uniform pulse in the uranium cube:
the spectral transient lasts about $0.1~\mu$s.}
\label{fig:cepulsed}
\end{figure}

\paragraph{The one-group prediction fails.}
Figure~\ref{fig:ce} shows that the simulated $Y(T)$ is much larger than the
one-group prediction, by factors of $1.7$ to $3.6$. The pulsed simulations
suggest the origin of the discrepancy (Fig.~\ref{fig:cepulsed}). The fission
rate per neutron decreases substantially during the first tens of
nanoseconds, while source and fission neutrons slow down below the fission
threshold of $^{238}$U through inelastic scattering. This spectral evolution
adds time scales that a one-speed model does not contain, as expected from
energy-dependent Feynman analyses~\citep{yamamoto2013energy}. The curves do
not establish that a single additional exponential would be sufficient.

At $10.5\%$, a BBM fit gives $\gamma_1=2.20\times10^6$~s$^{-1}$, compared
with the pulsed value $2.02\times10^6$~s$^{-1}$. Converting with the
one-group FREYA generation time gives a fitted quantity
$k_{\mathrm{1g}}=0.863$, compared with $0.874$ obtained from the one-group
rates; the point fit gives $0.768$. These reduced-model quantities must not
be identified with $k_{\mathrm{eval}}$, nor with an independently calculated
FREYA transport eigenvalue. At lower enrichments both fits degrade. This
example shows the limit of the one-speed reduction; it does not validate an
energy-dependent reactivity measurement.

\paragraph{Sensitivity to the fission model.}
The comparison of the fission algorithms is also instructive. At $3$ and
$7\%$, the analog FREYA runs give $Y$ values $16\%$ larger than standard
OpenMC. At $10.5\%$ the two coincide, although the sampling laws differ. The
standard runs use the evaluated mean multiplicity, which is larger than
FREYA's for fast fission of $^{238}$U (Supplementary Material, Table~S1).
A change in mean multiplication can offset a change in pair production, so
agreement of $Y$ alone is not a test of the collision algorithm. At $10.5\%$, the analog runs with the Zucker--Holden distribution
at the ENDF mean give $Y=70$, instead of $52.6$ with FREYA and $58.1$ with
the analog Bernoulli rounding.
Close to criticality the Feynman curve is very sensitive both to the first
moment of the multiplicity, through $k$, and to its second moment. The
$1.9\%$ difference between FREYA and the evaluation for fast fission of
$^{238}$U is not negligible for noise analysis.

\section{A spatial benchmark for inverse problems}
\label{sec:inversebenchmark}

The fits of Sec.~\ref{sec:k} assume that the leakage scale $a$ is known.
This section uses the cube to examine two extensions of that inverse problem. The
first shows that a pair of detectors can measure $a$ and $\rho$ separately.
The second extends the model to the third count moment, which connects it
to neutron multiplicity counting.

\subsection{Separating leakage and net multiplication}

\paragraph{Principle.}
Consider the two opposite faces of Fig.~\ref{fig:benchmark}. Let $I_R$ and
$I_L$ denote their counting currents, and form their sum and difference,
$I_\pm=I_R\pm I_L$. Let $R_\pm(\tau)$ be the autocovariance of $I_\pm$ at a
nonzero lag $\tau$; the delta contribution of each particle with itself at
zero lag is excluded. By reflection symmetry, the sum responds only to modes
with odd $n_x$ and the difference only to modes with even $n_x$. Integration
over a face requires odd $n_y$ and $n_z$. The slowest decay rates seen by
the two combinations are therefore those of the modes $(1,1,1)$ and
$(2,1,1)$,
\begin{equation}
 \gamma_+=\gamma_{111}=3a-\rho,\qquad
 \gamma_-=\gamma_{211}=6a-\rho,
 \qquad\text{hence}\qquad
 a=\frac{\gamma_--\gamma_+}{3},\quad \rho=\gamma_--2\gamma_+.
 \label{eq:parityinverse}
\end{equation}
The difference of the two rates removes the net multiplication and
determines the leakage scale. Neither covariance is mono-exponential at all
lags. Modal extraction and combinations of detector signals have
established precedents in coherence analysis, Rossi-$\alpha$ analysis and
pulsed-source
measurements~\citep{hashimoto1991,yamamoto2022rossi,yamamoto2023dmd,katano2019}.
Here they expose an ambiguity that matters for monitoring: two systems can
share the same $\gamma_+$ while having different leakage and multiplication.

\paragraph{Test.}
We constructed such a pair, in reduced units: $L=1$, $\lambda=4$, $S=2$,
$\gamma_+=0.6$, and $a=1$ (case A) or $a=1.3$ (case B). The source density
is proportional to the positive fundamental eigenfunction and both faces
have unit efficiency. The covariances $R_\pm$ are fitted with independent
free amplitudes, so that the source rate and the pair yield do not set
their scale; the source profile and the geometry remain prescribed. The
estimates of $a$ are (mean over independent records $\pm$ standard error of
the mean):
\begin{center}
\begin{tabular}{lcc}
\toprule
 & case A ($a=1$) & case B ($a=1.3$) \\
\midrule
Absorbing lattice (24 records) & $1.004\pm0.014$ & $1.291\pm0.025$ \\
Analog OpenMC (24 records) & $1.011\pm0.022$ & $1.308\pm0.028$ \\
\bottomrule
\end{tabular}
\end{center}
The standard deviations of the individual OpenMC estimates are $0.107$ and
$0.137$: a single record separates the two cases less sharply than
these means. If $\lambda$ is known independently,
\begin{equation}
 k=1-\frac{\gamma_+}{\lambda+\gamma_--\gamma_+}.
 \label{eq:parityk}
\end{equation}
The noise estimates of $k$ are $0.91401\pm0.00135$ and
$0.92473\pm0.00147$, compared with independent eigenvalue calculations
$0.91474\pm0.00064$ and $0.92722\pm0.00064$. Supplementary Material,
Sec.~S6 gives the spectra, protocols and convergence checks. Two
restrictions should be noted. The lattice is an approximation to the BBM,
and its exact discrete spectrum is used for validation. These OpenMC tests
use binary fissions, not FREYA.

\paragraph{Sensitivity to detector imbalance.}
The two detector responses must be balanced. A residual gain imbalance
$\delta$ gives
$I_-^{(\delta)}=(1+\delta)I_R-(1-\delta)I_L=I_-+\delta I_+$.
For a symmetric medium, source and ideal detector pair, the
cross-covariances between the two parities vanish, hence
\begin{equation}
 R_-^{(\delta)}(\tau)=R_-(\tau)+\delta^2R_+(\tau).
 \label{eq:imbalanceNSE}
\end{equation}
The contaminating symmetric term decays more slowly and eventually
dominates. It is also larger near criticality: at fixed source strength and
geometry, its leading amplitude diverges as $\gamma_+^{-2}$, whereas the
antisymmetric amplitude diverges as $\gamma_+^{-1}$. At fixed reduced lag
$a\tau$, keeping the contamination below a prescribed fraction thus
requires $|\delta|_{\max}=O(\sqrt{\gamma_+/a})$: the closer to critical, the
better the balance must be. In a test with $\gamma_+=0.12$ and $a=1$,
applying weights $1.05$ and $0.95$ and ignoring this imbalance gave
$\widehat a=0.862\pm0.036$, against $0.983\pm0.036$ for balanced channels
($\gamma_+$ was fixed to its true value to isolate the effect). Waiting for
a late tail does not help, since the desired antisymmetric signal decays
faster than its contaminant. Source asymmetry, finite detection efficiency
and detector time response require additional treatment.

\subsection{A spatial counterpart of the B\"ohnel construction}
\label{sec:boehnel}

The cube also connects time-correlation analysis to neutron multiplicity
counting, which uses the first three moments of the counts.

\paragraph{Equations.}
Let $M$ be the \emph{total} number of detected exits from one ancestor and
all its descendants, after the chain has died out, and let
$\Phi(\vx,z)=\E_{\vx}[z^{M}]$ be its generating function for an ancestor
starting at $\vx$.
It is the time-independent counterpart of $W_s$ in
Eq.~\eqref{eq:backward}, for a gate that covers the whole chain. If $\Gamma_d$ is the observed part of the
boundary, it satisfies
\begin{equation}
 D\nabla^2 \Phi+\lambda[G(\Phi)-\Phi]=0,\qquad
 \Phi|_{\Gamma_d}=1-\varepsilon+\varepsilon z,\quad
 \Phi|_{\partial V\setminus\Gamma_d}=1.
 \label{eq:spatialboehnel}
\end{equation}
Replacing spatial propagation by a uniform leakage rate $b$ gives
$(\lambda+b)\Phi=\lambda G(\Phi)+b(1-\varepsilon+\varepsilon z)$, the
corresponding point-model fixed-point equation.
Equation~\eqref{eq:spatialboehnel} thus retains the starting position and
the escape geometry in B\"ohnel's
construction~\citep{boehnel1985,prasad2012}. This specialization does not
supersede existing spatial third-order theories~\citep{endo2006third}.

Writing $u_r=\partial_z^r\Phi|_{z=1}$ for the factorial moments of $M$ and
$\nu_3=G^{(3)}(1)$, successive differentiation yields three linear
boundary-value problems,
\begin{align}
 \Lop u_1&=0,\\
 -\Lop u_2&=\lambda\nu_2u_1^2,\\
 -\Lop u_3&=3\lambda\nu_2u_1u_2+\lambda\nu_3u_1^3,
 \label{eq:threehierarchy}
\end{align}
with $u_1=\varepsilon$ on $\Gamma_d$ and zero on the unobserved boundary,
and $u_2=u_3=0$ on the whole boundary. When the whole boundary is observed,
$u_1=\varepsilon h$ and $u_2=\varepsilon^2\lambda\nu_2w$ in the notation of
Eq.~\eqref{eq:stationary}. Each order is a source term for the next, and in
a cube all three are solved with the same Dirichlet spectrum.

\paragraph{Observables.}
For Poisson immigration of single neutrons, the long-gate factorial
\emph{cumulant} rates of the counts are $c_r=S\langle u_r\rangle_p$, where
$\langle\cdot\rangle_p$ is the average over the source density. They give
\begin{equation}
 r_c=c_1,\qquad Y_\infty=c_2/c_1,\qquad X_\infty=c_3/c_1.
 \label{eq:momentoutputs}
\end{equation}
Here $r_c$ is the count rate and $X_\infty$ is the third Feynman moment,
the third-order analogue of $Y_\infty$. Raw factorial moments of stationary
gate counts also contain accidental
coincidences and must not be substituted for these connected cumulants.
Independent simulations of $2.4\times10^6$ chains verify all three moments
on the lattice. The third factorial chain moment is $2156\pm40$, against
the discrete prediction $2147.3$. Compound sources and grid convergence are
treated in Supplementary Material, Sec.~S7.

\paragraph{A reference map for inference.}
Houpert~\citep{houpert2024slides} discusses the inversion of count moments,
including the information provided by different gate durations. Lartaud,
Humbert and Garnier~\citep{lartaud2025} use point-model predictions as the
prior mean of Gaussian-process surrogates corrected by transport data. The
present hierarchy offers a spatial intermediate model for that role. For a
prescribed source profile, a known $\lambda$ and a specified family of
branching laws linking the offspring moments to $k$,
a useful reference map is
\begin{equation}
 (a,k,\varepsilon,S)\longmapsto
 (\gamma_+,\gamma_-,r_c,Y_\infty,X_\infty).
 \label{eq:forwardmap}
\end{equation}
For nonzero pair yield and detection efficiency, this map has a
constructive inverse. The two rates give $a$ and $k$
[Eqs.~\eqref{eq:parityinverse} and~\eqref{eq:parityk}]. Let
$U_r=\langle u_r\rangle_p$ be evaluated at unit efficiency. Then
$\varepsilon=Y_\infty U_1/U_2$ and $S=r_c/(\varepsilon U_1)$, while
$X_\infty=\varepsilon^2U_3/U_1$ provides a consistency check. We supply 72
synthetic design points and verify this noiseless inverse.

This algebraic test has a limited scope. It is not a validation of
posterior uncertainty, nor of a noisy inversion of all four parameters. A
surrogate cannot create identifiability when the source, the branching law
or the kinetic scale is left unconstrained. In transport applications,
observational covariance, surrogate uncertainty and diffusion-model
discrepancy must all be represented. Neutron--gamma inference, as studied
in Ref.~\citep{lartaud2025}, would require an additional particle type and
joint fission multiplicities, and is not implemented here.

\section{Discussion and conclusion}
\label{sec:conclusion}

The branching event that creates a spatial correlation also creates a pair
of possible detector counts. The two observables differ only in what
happens next: a count requires the descendant to reach the boundary within
the gate. Expressing that response through the Green's function leads to
Eq.~\eqref{eq:Ygeneral}, and the long-gate limit requires only the two
stationary problems of Eq.~\eqref{eq:stationary}. This connects the
confined-population formalism of Ref.~\citep{zoia2014clustering} directly
to spatial neutron-counting theory.

A spatial branching model, combined with a correct analog treatment of
fission, predicts boundary counting fluctuations and quantifies the bias of
the point approximation. In the one-speed cubes, point kinetics deviates
from transport at the longest simulated gate by $+48\%$, $+27\%$ and $-25\%$ at
nominal $k=0.9$, $0.8$ and $0.5$, respectively. The corresponding
absolute discrepancies of the spatial model are about $0.2\%$, $1\%$ and $10\%$. The
spatial model also reduces the bias on the inferred fundamental decay rate
and multiplication factor. This comparison is possible without fitting an
independent amplitude for each decay mode, because the modal weights are
determined by the geometry and the source.

This is a step towards using detector noise and numerical transport to
infer $\alpha$ eigenvalues and $k_{\mathrm{eff}}$, and its limits should be
stated plainly. The Feynman-curve inversion concerns the fundamental decay
rate with all other inputs prescribed; it does not recover an unknown
kinetic scale from one counting curve. The opposite-face benchmark adds a
second observable parity and separates leakage from net growth, under
symmetry and source assumptions. The evidence is limited to homogeneous
cubes, prompt neutrons and an ideal leakage detector. Finite-record
statistics, modal truncation and uncertainty in the material and source
must be included before the inversion is used as a measurement procedure.

The opposite-face example shows why geometry should be retained in a
forward model for inference. The three-moment hierarchy supplies a small
spatial reference problem in which parameter identifiability, approximation
error and surrogate error can be examined separately. Its role is to test
inference methods before they are applied to heterogeneous systems.

The uranium example shows that energy groups require a further extension.
The scalar mean operator becomes a matrix of diffusion, transfer and
production operators, and $\lambda\nu_2 f_s^2$ is replaced by a sum over
pairs of offspring groups, weighted by their joint factorial moments. The
common-ancestor construction is preserved, but an average multiplicity
alone is no longer sufficient. This is a natural way to include the
spectral evolution seen in the uranium cube and, with precursor types,
delayed emission. Source fission multiplets supply the additional terms
listed in Supplementary Material, Sec.~S7. These extensions are necessary
for fuel debris applications; they cannot be inferred from the present
one-speed comparisons.

\paragraph{Code and data availability.}
The OpenMC modifications, analysis scripts and stored block statistics are
identified by source snapshot in Supplementary Material, Sec.~S5 and are
available from the authors. The stored block statistics permit the reported
fit-sensitivity analysis; raw detection-time records were not retained for
those earlier runs. The opposite-face study includes its counting records,
OpenMC surface histories, seeds and analysis scripts, with independent
binary-fission transport runs and chain-moment tests. Reproducing the
FREYA transport runs also requires the specified nuclear data and the LLNL
Fission library. The new binary-fission tests do not require that library.

\bibliographystyle{unsrtnat}
\bibliography{refs}

\end{document}


\maketitle
\section{Analog implementation and multiplicity checks}
\subsection{Coupling with the LLNL Fission library and FREYA}

The Fission library v2.0.5 was downloaded from LLNL and compiled with its
FREYA Fortran component. Only the uranium data were kept: the FREYA tables
of the compound nuclei $^{234,236,239}$U (induced fission of $^{233,235,238}$U)
and the spontaneous fission of $^{238}$U, with the reaction list
\texttt{react.dat} restricted accordingly. OpenMC is linked to the library
when configured with \texttt{-DOPENMC\_USE\_FREYA=ON
-DFREYA\_ROOT=\ldots}; without this option only the \texttt{bernoulli} model
is available. Three points of the coupling deserve attention.
\begin{enumerate}
  \item The multiplicity is taken from the event (\texttt{getnnu\_}), rather than
  only the energies and directions of neutrons whose number was drawn by
  OpenMC. Otherwise $P(\nu)$ would remain the two-point distribution of
  the rounding law in the main text.
  \item The random numbers of the library are drawn from the stream of the
  particle being transported, through a thread-local pointer passed to the
  callback registered with \texttt{setrngd\_}. For the verification configuration, two runs with the same seed produce
  identical multiplicity histograms. Reproducibility across thread counts
  has not been assessed. Calls
  to the library, which keeps the event in global state, are serialized in an
  OpenMP critical section.
  \item The energies returned by FREYA are checked: a neutron with a
  non-finite energy, or an energy outside the range of the cross-section data,
  is resampled from the evaluated fission spectrum. Clamping such energies
  to the upper bound of the data instead leads to out-of-bounds
  reads of the logarithmic energy grid. About
  one neutron in $10^6$ is concerned.
\end{enumerate}
Multigroup mode is supported as well. The multiplicity is then that of a
nuclide and incident energy given by \texttt{analog\_fission\_mg\_za} and
\texttt{analog\_fission\_mg\_energy}, and neutrons are emitted isotropically
in the fission spectrum of the material. A histogram of the prompt
multiplicities, sorted by nuclide and incident energy, is written at the end
of the run together with the numbers of sampled and expected
($\sum\sigma_f/\sigma_t$) fissions.

\subsection{Verification}

\begin{figure}[t]
\centering
\includegraphics[width=\textwidth]{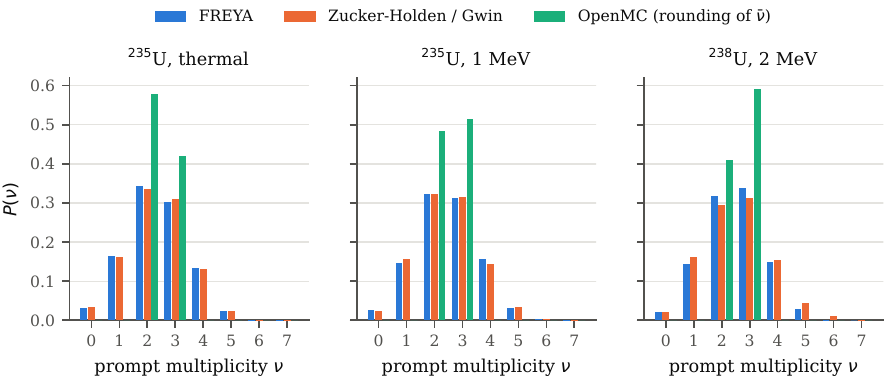}
\caption{Prompt multiplicity distributions of FREYA, of the Zucker--Holden
and Gwin data, and of the rounding of $\nub$ used by standard OpenMC.}
\label{fig:pnu}
\end{figure}

On a bare natural-uranium sphere irradiated by $2$~MeV neutrons, the number
of sampled fissions agrees with $\sum\sigma_f/\sigma_t$ within one standard
deviation for the three models (e.g. $3363$ sampled for $3361.9$ expected
with FREYA). Table~\ref{tab:pnu} compares the multiplicity distributions,
sampled with $10^6$ events, and Fig.~\ref{fig:pnu} displays them. The
Zucker--Holden and Gwin distributions, used at the evaluated mean, give Diven
factors $D_2\simeq0.79$--$0.83$. FREYA reproduces the thermal $^{235}$U values
within $0.2\%$ for $\nub$ and $D_2$, but its mean multiplicity for fast
fission of $^{238}$U at $2$~MeV is $1.9\%$ below ENDF/B-VII.1 and its second
factorial moment $9\%$ below the LLNL tabulated distribution. OpenMC's rounding
underestimates $\langle\nu(\nu-1)\rangle$ by $20$--$25\%$ and $\langle\nu(\nu-1)(\nu-2)
\rangle$ by more than $60\%$.

\begin{table}[t]
\centering
\caption{Prompt multiplicity moments for three reactions. $\nub_p$ is the
ENDF/B-VII.1 value used by OpenMC. $\nu_2=\langle\nu(\nu-1)\rangle$,
$\nu_3=\langle\nu(\nu-1)(\nu-2)\rangle$, $D_2=\nu_2/\nub^2$.}
\label{tab:pnu}
\small
\begin{tabular}{llccccc}
\toprule
Reaction & Model & $\nub_p$ (ENDF) & $\nub$ & $\nu_2$ & $\nu_3$ & $D_2$ \\
\midrule
$^{235}$U, thermal & FREYA & 2.421 & 2.416 & 4.619 & 6.658 & 0.791 \\
 & Zucker--Holden/Gwin & & 2.423 & 4.652 & 6.704 & 0.793 \\
 & OpenMC rounding & & 2.421 & 3.683 & 2.525 & 0.629 \\
\midrule
$^{235}$U, 1 MeV & FREYA & 2.516 & 2.533 & 5.135 & 7.980 & 0.801 \\
 & Zucker--Holden & & 2.516 & 5.069 & 7.911 & 0.801 \\
 & OpenMC rounding & & 2.516 & 4.064 & 3.096 & 0.642 \\
\midrule
$^{238}$U, 2 MeV & FREYA & 2.591 & 2.541 & 5.062 & 7.477 & 0.784 \\
 & Zucker--Holden & & 2.591 & 5.537 & 9.678 & 0.825 \\
 & OpenMC rounding & & 2.591 & 4.363 & 3.544 & 0.650 \\
\bottomrule
\end{tabular}
\end{table}

These checks concern multiplicity and fission selection. They do not validate
joint energy--angle distributions or delayed-neutron correlations. In the
one-speed calculations only the multiplicity law is taken from FREYA;
the emission spectrum is the prescribed material spectrum and directions are
isotropic. Survival biasing and weight windows are rejected in analog mode.
The calculations considered here use unit-weight sources in fixed-source mode.
The optional multiplicity histogram is written by the master MPI rank in the
supplied implementation; it is not an MPI-reduced diagnostic.

\paragraph{Multiplicity options.}
\begin{description}
  \item[\texttt{bernoulli}] OpenMC's rounding of $\nub$, the rounded-multiplicity rule
  with $x=\nub$. Combined with the analog collision it isolates the effect of
  the multiplicity distribution.
  \item[\texttt{llnl}] the tabulated $P(\nu)$ of the LLNL Fission
  library~\citep{verbeke2016fission}, based on Zucker and
  Holden~\citep{zucker1986} and Gwin et al.~\citep{gwin1984}. The selected
  option mixes tabulated distributions to match the evaluated $\nub_p(E)$,
  with a Terrell-law fallback outside the tabulated range.
  \item[\texttt{freya}] complete fission events generated by
  FREYA~\citep{verbeke2018correlated,vogt2009freya}: the prompt multiplicity
  and the correlated energies and directions of the neutrons are taken from
  the same event. FREYA computes $\nub$ from its own physics model and ignores
  the evaluated value.
\end{description}

\section{Separable evaluation and analytical checks}
For a separable source $p(\vx)=\prod_{i=1}^3p_i(x_i)$, write
$s_n^{(i)}=\int_{-L}^L p_i(x)\psi_n(x)\,\dd x$. Then
\begin{equation}
 A_{\bm k\bm l}=\int_V\widehat c\,\varphi_{\bm k}\varphi_{\bm l}
 =\int_0^\infty e^{\rho u}\prod_{i=1}^3 B_{k_il_i}^{(i)}(u)\,\dd u,
 \qquad B_{kl}^{(i)}(u)=\sum_n s_n^{(i)}\tau_{nkl}e^{-an^2u}.
\end{equation}
Here $1/\gamma_{\vn}=\int_0^\infty e^{-\gamma_{\vn}u}\dd u$ separates the
three coordinates. For the centered uniform source only odd indices
contribute; $s_n=c_n/(2L)$ if the source fills the absorbing cube.
For a source restricted to $[-L_s,L_s]^3$, the same coefficients are obtained
by integrating over $[-L_s,L_s]$, with normalization $1/(2L_s)$.
This distinction matters in transport comparisons: the source occupies the
physical cube, while the absorbing diffusion boundary is extrapolated.

Let $w_{\bm k}=C_{\bm k}b_{\bm k}$. Symmetry of $A$ and of the gate kernel
reduces the numerical evaluation to
\begin{equation}
 Y(T)=\frac{\varepsilon\lambda\nu_2}{\langle h\rangle}
 \sum_{\bm k} d_{\bm k}H(\gamma_{\bm k}T),\qquad
 d_{\bm k}=\frac{2w_{\bm k}}{\gamma_{\bm k}}
 \sum_{\bm l}\frac{A_{\bm k\bm l}w_{\bm l}}
 {\gamma_{\bm k}+\gamma_{\bm l}}.
 \label{eq:fast}
\end{equation}
This rearrangement avoids constructing a pair-mode matrix for every gate.
For a small truncation ($m=3$), its values agree with the original direct
pair-mode evaluation to a relative difference below $4\times10^{-15}$ on
ten logarithmically spaced gates in the one-speed $k=0.8$ case.

For completeness, splitting ancestors between $s<0$ and $0<s<T$ gives
\begin{align}
 F(g,h;T)&=\frac{1}{gh}\left[T-\frac{1-e^{-gT}}{g}
 -\frac{1-e^{-hT}}{h}+\frac{1-e^{-(g+h)T}}{g+h}\right]\\
 &\quad+\frac{(1-e^{-gT})(1-e^{-hT})}{gh(g+h)}
 =\frac{T}{gh}\frac{hH(gT)+gH(hT)}{g+h}.
\end{align}
The second term is essential: omitting ancestors born before the gate
would describe a transient experiment. The limits are
$F(g,g;T)=TH(gT)/g^2$, $F\sim T^2/(g+h)$ as $T\to0$,
and $F\sim T/(gh)$ as $T\to\infty$. The last two limits hold mode by mode;
interchanging a short-time limit with the infinite boundary-flux sum
requires a separate convergence argument.

The implementation evaluates the source-mode sum with 25 odd
indices per coordinate (or more if $m>25$) and the Laplace integral on a
600-point logarithmic grid, plus its origin. Its mean-leakage calculation
uses 1000 odd modes. The displayed forward curves extrapolate the $m=11$
and $m=15$ results linearly in $1/m$. This is a numerical prescription,
not an error bound. A converged forward or inverse calculation must also
vary the source-mode cutoff and quadrature, not only $m$.

\section{Sensitivity of the numerical comparisons}
\subsection{Gate covariance and modal truncation}
The stored one-speed data contain four independently seeded runs, each
partitioned into eight consecutive segments. The resulting 32 vectors
$Y_b(T_i)$ allow the covariance of their mean to be estimated as
\begin{equation}
 \widehat C_{ij}=\frac{1}{32(32-1)}\sum_{b=1}^{32}
 [Y_b(T_i)-\bar Y(T_i)][Y_b(T_j)-\bar Y(T_j)].
\end{equation}
This treats segments as approximately independent; only four complete
records are independently seeded. For the 16 gates above 10 ns, a free
positive amplitude is profiled out analytically at each trial decay rate,
and the remaining one-dimensional objective is minimized. Diagonal fits
use $\sigma_i=\max(\operatorname{SE}_i,10^{-3}\bar Y_i)$, as in the original
analysis. Covariance-weighted fits use the full empirical matrix without
regularization. Their matrix condition numbers are approximately
$2.8\times10^3$, $1.5\times10^4$ and $1.1\times10^5$; these fits are a
sensitivity diagnostic, not a basis for precise confidence intervals.

\begin{table}[ht]
\centering
\caption{Reanalysis of the stored one-speed FREYA block statistics.
The entries are inferred $k$, with all material and source inputs fixed.}
\begin{tabular}{llccc}
\toprule
Weighting & $m$ & nominal $k=0.5$ & $0.8$ & $0.9$\\
\midrule
Diagonal & 5 & 0.4726 & 0.8073 & 0.9047 \\
Diagonal & 7 & 0.5110 & 0.8142 & 0.9058 \\
Diagonal & 9 & 0.5225 & 0.8151 & 0.9056 \\
Full covariance & 5 & 0.4687 & 0.8070 & 0.9054 \\
Full covariance & 7 & 0.5167 & 0.8192 & 0.9082 \\
Full covariance & 9 & 0.5317 & 0.8211 & 0.9084 \\
\bottomrule
\end{tabular}
\end{table}
The $m=7$ diagonal results reproduce the original estimates. Increasing
$m$ does not uniformly improve agreement with the transport reference:
truncation can compensate for diffusion error. At $m=9$, the diagonal
weighted residual sums are $3.6$, $55.2$, $149.8$
for 16 observations and two fitted parameters. Thus even visually close
curves are not generally consistent with the reported sampling errors
alone. Neither this calculation nor a change of weighting removes the
need for convergence and model-error assessment.

\subsection{Finite-step BBM simulations}
Table~\ref{tab:dt} compares the three available time steps. The middle step
has $8\times10^5$ histories; the other two have $4\times10^5$. Uncertainties
are the stored standard errors. The largest gate is the same within each
configuration. The records do not demonstrate a stable extrapolation in
$\dd t$, especially at $k=0.95$. Differences between steps include sampling
variation. Leakage is assigned to the midpoint of a step, reproduction is
split from diffusion, and the product of single-face bridge survival
probabilities is an approximation for opposite faces. These are additional
reasons not to describe the simulation as exact.
\begin{table}[ht]
\centering
\caption{Stored mean leaks per chain and long-gate $Y$ for the BBM runs.}
\label{tab:dt}
\begin{tabular}{cccc}
\toprule
$k$ & $\dd t$ & mean leaks & $Y(T_{\max})\pm\operatorname{SE}$\\
\midrule
0.5 & $0.00020$ & 0.9120 & $0.932\pm0.016$ \\
0.5 & $0.00010$ & 0.9094 & $0.903\pm0.019$ \\
0.5 & $0.00005$ & 0.9099 & $0.881\pm0.014$ \\
0.8 & $0.00020$ & 1.6382 & $11.271\pm0.221$ \\
0.8 & $0.00010$ & 1.6386 & $11.343\pm0.162$ \\
0.8 & $0.00005$ & 1.6475 & $11.569\pm0.538$ \\
0.95 & $0.00020$ & 5.0245 & $284.928\pm11.136$ \\
0.95 & $0.00010$ & 5.0152 & $294.663\pm8.384$ \\
0.95 & $0.00005$ & 5.1041 & $318.490\pm13.179$ \\
\bottomrule
\end{tabular}
\end{table}

\subsection{Conditioning on the number of source histories}
Uniformly distributing a fixed number $N$ of independent histories over an
interval of length $T_m$ is a Poisson source conditioned on $N$. If $Z$ is
the count contribution of one such randomly timed history to a specified
gate, then $C=\sum_{j=1}^N Z_j$ and
\begin{equation}
 Y_N=\frac{\E[Z(Z-1)]}{\E[Z]}-\E[Z].
\end{equation}
For a Poisson number of histories of mean $N$ in the same interval, the last
term is absent. Away from record edges,
$\E[Z]\simeq\varepsilon\langle h\rangle T/T_m$. For the one-speed records,
the nominal count rate is $2\gamma_1$, giving a leading correction
$2\gamma_1T/N$. At the largest stored gate, its values are
$2.2\times10^{-4}$, $8.8\times10^{-5}$ and $4.4\times10^{-5}$
for nominal $k=0.5,0.8,0.9$, respectively. This ensemble correction is small
in these records. It does not quantify the bias or uncertainty of the
sample-variance estimator over finite, correlated gates, nor the discarded
start-up interval. Raw time records would be needed to vary block length
and gate placement independently.

\section{Population moments}

Another observable is the number $n_V(t)$ of particles
present in the cube at time $t$, for $N_0$ particles placed uniformly at
$t=0$. The same formalism gives exact expressions for this observable.
Averaging the mean $m_t(\vx)$ and the factorial moment
$\lambda\nu_2\int_0^t\dd s\int\mathcal{G}_{t-s}m_s^2$~\citep{zoia2014clustering}
over the initial position, and using the factorization
$m_t(\vx)=e^{\rho t}\prod_iu(t,x_i)$, where $u$ solves the one-dimensional
heat equation from $u(0,x)=1$ with Dirichlet conditions, we find
\begin{equation}
  \langle m_1\rangle(t)=e^{\rho t}\bar u(t)^3,\qquad
  \langle m_2\rangle(t)=\frac{\lambda\nu_2}{|V|}\int_0^t\dd s\,
  e^{\rho(t+s)}\left[\int_{-L}^{L}u(t-s,x)\,u(s,x)^2\,\dd x\right]^3,
  \label{eq:population}
\end{equation}
and $\Var[n_V]/\E[n_V]=1+\langle m_2\rangle/\langle m_1\rangle-\langle m_1\rangle$.
In the fundamental-mode approximation this reads
$1+\kappa\lambda\nu_2(1-e^{-\gamma_1t})/\gamma_1-(8/\pi^2)^3e^{-\gamma_1t}$.

\begin{figure}[t]
\centering
\includegraphics[width=0.72\textwidth]{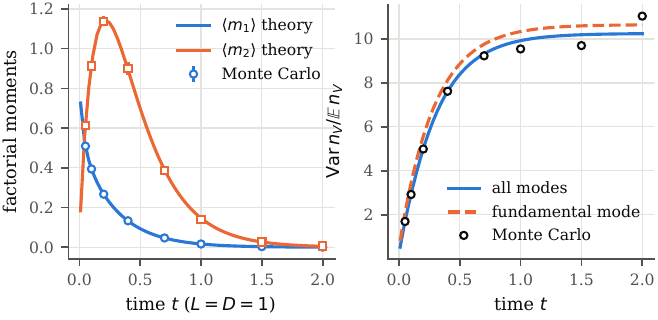}
\caption{Population in the cube for particles started uniformly ($k=0.8$):
factorial moments per initial particle, and variance-to-mean ratio of $N_0$
particles, Eq.~\eqref{eq:population}, versus the finite-step simulation.}
\label{fig:population}
\end{figure}
Here $N_0$ is fixed and initial positions are independent. The subtraction
of $\langle m_1\rangle$ is consequently required. A Poisson initial
population has a different variance-to-mean ratio. This comparison concerns
the population inside the domain and is separate from boundary counting.

\section{Source snapshots and reproducibility}
The source snapshots used for this work are:
\begin{itemize}
\item Analysis and stored results, \texttt{bbm-neutron-noise}:\\
\texttt{484bf9028cb13092c3b83f10beed9848f9db82a0}.
\item Modified OpenMC, branch \texttt{analog-fission-freya}:\\
\texttt{6748a601b5456b9a02352dfebb23cc8895240a4e}.
\end{itemize}
The repositories are:\\
{\small\url{https://github.com/bessemoa/bbm-neutron-noise}}\\
{\small\url{https://github.com/bessemoa/openmc-analog-fission}}\\
Access can be requested from the authors. The calculations use the LLNL Fission library
2.0.5, whose available technical manual describes release
2.0.2~\citep{verbeke2016fission,llnllibrary}, and the evaluated data specified
in the main text.

The reanalysis uses stored segment estimates in \texttt{sim/data/mg\_noise},
BBM summaries in \texttt{sim/data/bbm\_mc\_*.npz}, and
\texttt{bbm\_theory.py}. No new OpenMC histories were generated in this
reanalysis. The verification package includes the inputs used for the new
sensitivity tables and their postprocessing script. Original detection-time
records were not retained: the simulation scripts delete their temporary
surface-source files. The fit-sensitivity calculations reuse stored
statistics and do not constitute independent transport runs.

\section{Opposite-face covariance benchmark}
\label{sec:newbenchmark}
The additional calculations use the same OpenMC snapshot as Sec.~S5,
compiled without the LLNL Fission library. The analog \texttt{bernoulli}
option is used with induced multiplicity exactly two. These are new
transport histories, distinct from the stored FREYA results discussed above.
All raw surface records of these new runs have been retained.

With source density $p=\varphi_{111}/C_1$, $C_1=\int_V\varphi_{111}$,
the stationary concentration per unit source is
$\widehat c=\varphi_{111}/(\gamma_+C_1)$.
Define $d_{\pm,\mathbf n}=-D\int_{\partial V}w_\pm\partial_n\varphi_{\mathbf n}$,
where $w_+$ is one on both observed faces, $w_-$ is one on the right and
minus one on the left, and both vanish elsewhere. For unit efficiency,
\begin{align}
R_\pm(\tau)&=\sum_{\mathbf n}A_{\pm,\mathbf n}e^{-\gamma_{\mathbf n}\tau},\\
A_{\pm,\mathbf n}&=\frac{S\lambda\nu_2}{\gamma_+C_1}
d_{\pm,\mathbf n}\sum_{\mathbf m}
\frac{d_{\pm,\mathbf m}\int_V\varphi_{111}\varphi_{\mathbf m}\varphi_{\mathbf n}}
{\gamma_{\mathbf m}+\gamma_{\mathbf n}}.
\end{align}
The index sets are restricted by detector parity. Coefficients need not
all be positive. At fixed geometry and nonzero pair yield, the leading
symmetric amplitude is $O(\gamma_+^{-2})$ because the term
$\mathbf m=\mathbf n=(1,1,1)$ has denominator $2\gamma_+$ in addition to
the stationary factor $1/\gamma_+$. No such extra singularity exists in
the antisymmetric sector, whose amplitude is $O(\gamma_+^{-1})$.
The exact imbalance criterion for a contamination fraction $q$ at lag
$\tau$ is $|\delta|\le\sqrt{qR_-(\tau)/R_+(\tau)}$.
Numerical slopes versus $\gamma_+/a$ are -2.0009, -1.0016 and 0.4997 for
the two amplitudes and the tolerance, respectively. This is a deterministic
asymptotic check, not a Monte Carlo measurement of the critical exponent.

The absorbing random walk uses $n$ interior nodes per coordinate,
$h=2/(n+1)$ and jump rate $D/h^2$ to each neighbour. Reaction times are
exponential at rate $\lambda=4$, with $q_2=(1+\rho_h/\lambda)/2$ and
$q_0=1-q_2$. Its exact eigenvalues are
$b_j^h=4Dh^{-2}\sin^2[j\pi/(2(n+1))]$. We set
$\rho_h=3b_1^h-\gamma_+$ and use this discrete spectrum for comparison
and inversion. The continuum estimate of $a$ would be low by 1.60\%
at $n=15$ and 0.40\% at $n=31$. The Monte Carlo is exact for the lattice,
not for continuous Brownian first passage.

All source arrivals are Poisson with rate $S=2$. We discard a build-up of
$30/\gamma_+$ and bin the two face counts at width 0.025. A and B have
$\gamma_+=0.6$, $a=1$ and $1.3$, with 24 independent records of length
20000 on $n=15$. The near-critical case has $a=1$, $\gamma_+=0.12$,
16 records of length 12000. The $n=31$ control uses 12 records of length
12000 at the A parameters. Covariances are computed from counting records,
without genealogy. Fits use separate free amplitudes and fixed lag windows
$[0.1,4]$ and $[0.1,1.2]$ for sum and difference. Uncertainty is assessed
between complete independent records, not by treating lags as independent.
The covariance model includes averaging over each pair of bins.

The one-speed OpenMC cases have $D=4a/\pi^2$,
$v=3D\Sigma_t$, $\Sigma_a=4/v$, $\Sigma_f=(4+\rho)/(2v)$,
and $\Sigma_s=\Sigma_t-\Sigma_a$. Their physical half-side is
$1-0.7104/\Sigma_t$. The cosine source profile is restricted to the
physical cube and tabulated with 1001 points per coordinate. Each run has
a Poisson-distributed number of roots and independently uniform source
times, ensuring Poisson immigration. At $\Sigma_t=20$, A and B each use
24 records of length 10000. A control at $\Sigma_t=40$ uses 16 records.
All surface weights were unity and the surface-bank capacity was not reached.
Separate eigenvalue runs use 100 active generations of 10000 particles,
after 40 inactive generations. Eigenvalue population control is not used
to generate the noise records.

\begin{table}[ht]\centering\small
\begin{tabular}{lrrr}\toprule
Case & $\widehat a$ & $\widehat\gamma_+$ & SD of $\widehat a$\\\midrule
Lattice A, $n=15$ & $1.004\pm0.014$ & $0.608\pm0.007$ & 0.067\\
Lattice B, $n=15$ & $1.291\pm0.025$ & $0.597\pm0.006$ & 0.125\\
Lattice A, $n=31$ & $0.998\pm0.027$ & $0.584\pm0.008$ & 0.092\\
OpenMC A, $\Sigma_t=20$ & $1.011\pm0.022$ & $0.604\pm0.009$ & 0.107\\
OpenMC B, $\Sigma_t=20$ & $1.308\pm0.028$ & $0.595\pm0.009$ & 0.137\\
OpenMC A, $\Sigma_t=40$ & $1.025\pm0.019$ & $0.600\pm0.008$ & 0.075\\\bottomrule
\end{tabular}
\caption{Mean estimates and standard errors of these means. SD denotes
the standard deviation across records, not the error of the mean.}
\end{table}
The initial eight-record OpenMC A pilot gave $\widehat a=1.127\pm0.033$,
whereas the full 24-record result is compatible with the target. Late-window
fits are less stable. Raising the lower lag from 0.1 to 0.4 gives
$1.628\pm0.116$ for OpenMC B. This loss of stability is retained as a
limitation, not removed by selecting a favourable window.
An independent Lyapunov-equation solution on a small lattice verifies
the modal covariance to $5.6\times10^{-15}$ relative accuracy. In the
continuum series, increasing the mode limit from 15 to 31 changes the
tested covariances by less than 0.35\% at lags at least 0.1.

\begin{figure}[ht]\centering
\includegraphics[width=\linewidth]{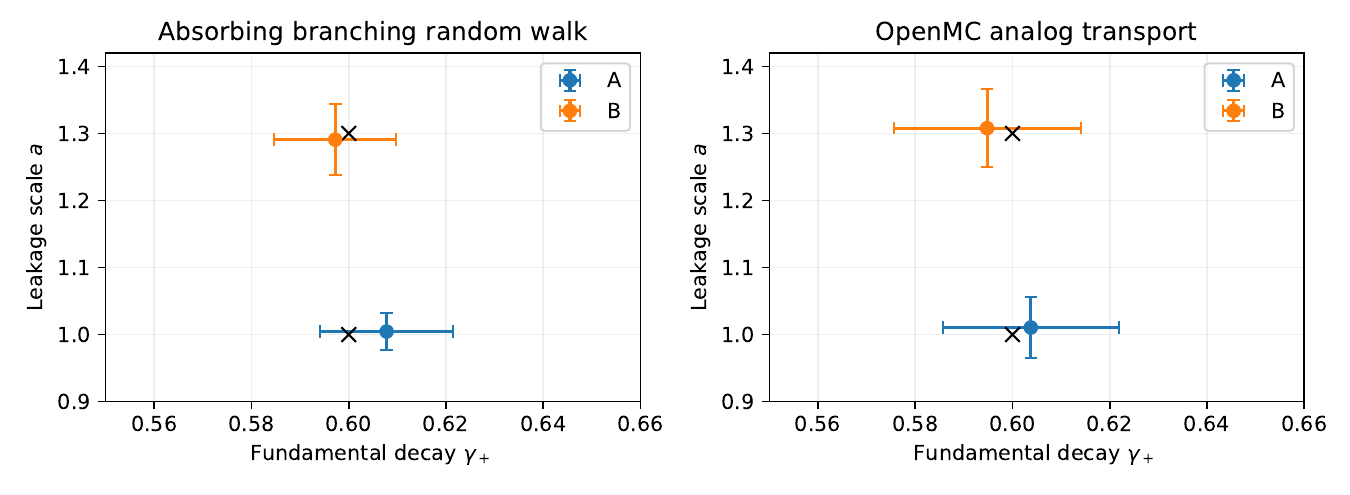}
\caption{Leakage identification from two parity channels. Crosses are
prescribed model parameters. Bars are 95\% Student intervals for the
mean estimates, not prediction intervals for a single acquisition.}
\end{figure}
\begin{figure}[ht]\centering
\includegraphics[width=\linewidth]{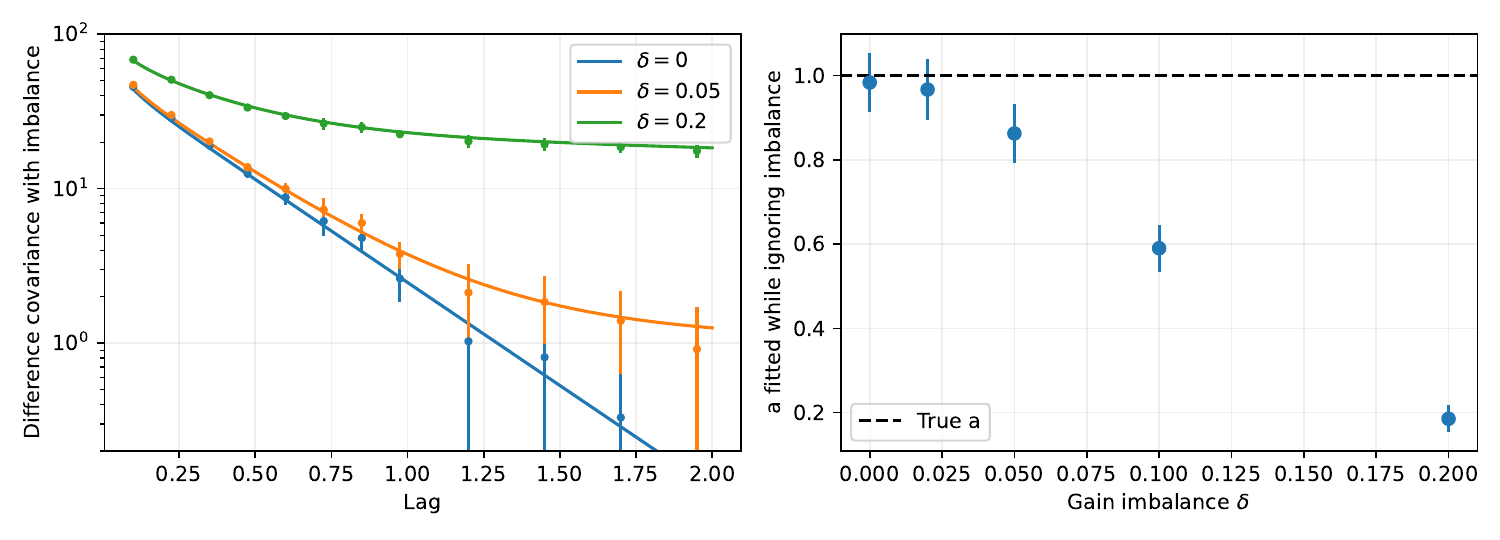}
\caption{Near-critical lattice case, $a=1$, $\gamma_+=0.12$.
Left: theory and Monte Carlo covariances after gain weighting.
Right: bias when imbalance is ignored, with $\gamma_+$ fixed to its true
value and the difference fitted on $[0.3,1.2]$. Points use the same records
for every imbalance and are therefore paired.}
\end{figure}

\section{Spatial multiplicity hierarchy and inverse reference map}
For an observed boundary $\Gamma_d$, the hierarchy in the main text is
obtained by differentiating its stationary generating-function equation.
It holds for a subcritical mean operator and finite third offspring moment.
The first boundary condition is affine in $z$, so all higher factorial
moments have zero boundary data. This explains why the same linear
resolvent can be reused for every order.

For source events emitting a batch with generating function $Q$, let
$s_j=Q^{(j)}(1)$. All neutrons in a batch start at its location and evolve
independently thereafter. The long-gate factorial cumulant rates become
\begin{align}
c_1&=S\langle s_1u_1\rangle_p,\\
c_2&=S\langle s_1u_2+s_2u_1^2\rangle_p,\\
c_3&=S\langle s_1u_3+3s_2u_1u_2+s_3u_1^3\rangle_p.
\end{align}
They follow from the composition $Q(\Phi)$, where $\Phi$ is the chain generating function of the main text, and independent Poisson source
events. More precisely,
$c_r=\lim_{T\to\infty}T^{-1}\partial_z^r\log\E[z^{N(T)}]|_{z=1}$.
Thus the binomial cumulant rates are $c_r/r!$. These quantities remove
accidental coincidences and should not be confused with raw factorial
gate moments. A spatially distributed batch would require a joint source
kernel instead of the local powers above. Only $Q(z)=z$ is tested here.

For the binary lattice benchmark ($a=1$, $\gamma_+=0.6$, $\lambda=4$),
we solved the three Poisson problems by discrete sine transforms. A
separate Monte Carlo sampled 24 batches of 100000 independent complete
chains, recording the total exits through the two observed faces. This
uses genealogy to verify the direct moment calculation, not to estimate
the achievable uncertainty of an experimental gate-moment measurement.
\begin{table}[ht]\centering
\begin{tabular}{lrr}\toprule
Chain factorial moment & Exact lattice, $n=15$ & Monte Carlo\\\midrule
$\langle u_1\rangle_p$ & 1.66132 & $1.65941\pm0.00414$\\
$\langle u_2\rangle_p$ & 33.9590 & $33.9515\pm0.3214$\\
$\langle u_3\rangle_p$ & 2147.32 & $2156.14\pm39.52$\\\bottomrule
\end{tabular}
\caption{Errors are standard errors across 24 independent batches.}
\end{table}
The second moment also equals $2\int_0^\infty R_+(\tau)\,\dd\tau/S$,
verified independently to $1.3\times10^{-15}$ relative accuracy. The
three moments at $n=31,63,127$ are respectively
$(1.66533,34.3528,2193.09)$, $(1.66633,34.4524,2204.74)$ and
$(1.66658,34.4774,2207.67)$. The third moment changes by about 0.13\%
between the two finest grids. The $n=15$ result must not be reported as
the continuum value.

The supplied reference dataset has 72 points, with
$a\in\{0.8,1,1.2\}$, $k\in\{0.7,0.8,0.9,0.94\}$,
$\varepsilon\in\{0.2,0.5,1\}$ and $S\in\{1,2\}$. It uses $n=31$,
known $\lambda=4$, binary reproduction and the same source profile.
Its inputs control the discrete multiplication factor through
$\gamma_+=(1-k)(\lambda+3b_1^h)$. The direct and inverse maps consistently
use that discrete spectrum. Noiseless recovery has maximum relative
error $1.2\times10^{-15}$. The smallest singular value of the relative
Jacobian of $(\gamma_+,\gamma_-,r_c,Y_\infty)$ with respect to
$(a,k,\varepsilon,S)$ is at least 0.198 on these points.
This is a numerical check on the specified design, not a universal
conditioning bound or an uncertainty-coverage result.

In the continuum, the two rates determine $a,\rho$ and, given $\lambda$,
$k$. Independent thinning implies $u_r(\varepsilon)=\varepsilon^r u_r(1)$,
which gives the constructive inverse in the main text. $X_\infty$ is an
extra model check rather than an additional unknown. If $\nu_2$, source
batch statistics or the source profile are unknown, this inversion no
longer follows without additional information.

For surrogate-assisted inference, the intended architecture is
$f_{\mathrm{transport}}(\theta)=f_{\mathrm{cube}}(\theta)+d(\theta)$,
with a trained correction $d$. The count rate and correlation moments
share data and must have a joint observational covariance. Surrogate
predictive uncertainty is distinct from unresolved physical-model
discrepancy. No Gaussian process or Bayesian posterior is trained in this
benchmark. The independently checked map provides a reference for testing
such procedures, in the spirit of Refs.~\citep{houpert2024slides,lartaud2025}.

\bibliographystyle{unsrtnat}
\bibliography{refs}